\documentclass[conference]{IEEEtran}
\IEEEoverridecommandlockouts

\usepackage{graphicx}
\usepackage{subcaption}
\usepackage{amsmath}
\usepackage{color}
\usepackage{algorithmic}
\usepackage[ruled]{algorithm2e}
\usepackage{cite}
\usepackage{float}
\usepackage{mathrsfs} 
\usepackage{booktabs}
\usepackage{amssymb}
\usepackage{lipsum} 
\usepackage{cuted}
\usepackage{flushend}
\usepackage{stfloats}
\usepackage{hyperref}

\newtheorem{theorem}{\textbf{Theorem}}
\newtheorem{corollary}{Corollary}

\newtheorem{lemma}{\textbf{Lemma}}

\newtheorem{proposition}{\textbf{Proposition}}

\newtheorem{example}{Example}
\newtheorem{definition}{\textbf{Definition}}

\def\BibTeX{{\rm B\kern-.05em{\sc i\kern-.025em b}\kern-.08em
    T\kern-.1667em\lower.7ex\hbox{E}\kern-.125emX}}
\begin{document}

\title{Uncertainty-of-Information Scheduling: A Restless Multi-armed Bandit Framework
}

\author{Gongpu~Chen,
	Soung Chang Liew,~\IEEEmembership{Fellow,~IEEE,}
	Yulin~Shao
	
	\thanks{Manuscript received February 06, 2021; revised November 13, 2021; accepted May 17, 2022. This work was supported in part by the General Research Funds (Project No. 14200221) established under the University Grant Committee of the Hong Kong Special Administrative Region, China. (\textit{Corresponding author: Soung Chang Liew.}) }
	
	\thanks{G. Chen and S.C. Liew are with the Department of Information Engineering, The Chinese University of Hong Kong, Shatin, Hong Kong (e-mail: \{gpchen, soung\}@ie.cuhk.edu.hk).}
	
	\thanks{Y. Shao was with The Chinese University of Hong Kong. He is now with Imperial College London, London SW7 2AZ, U.K. (e-mail:
		y.shao@imperial.ac.uk).}
	
}

\maketitle

\begin{abstract}
This paper proposes using the uncertainty of information (UoI), measured by Shannon's entropy, as a metric for information freshness. We consider a system in which a central monitor observes $M$ binary Markov processes through $m$ communication channels ($m<M$). The UoI of a Markov process corresponds to the monitor's uncertainty about its state. At each time step, only $m$ Markov processes can be selected to update their states to the monitor; hence there is a tradeoff among the UoIs of the processes that depend on the scheduling policy used to select the processes to be updated. The age of information (AoI) of a process corresponds to the time since its last update. In general, the associated UoI can be a non-increasing function, or even an oscillating function, of its AoI, making the scheduling problem particularly challenging.  This paper investigates scheduling policies that aim to minimize the average sum-UoI of the processes over the infinite time horizon. We formulate the problem as a restless multi-armed bandit (RMAB) problem, and  develop a Whittle index policy that is near-optimal for the RMAB after proving its indexability. We further provide an iterative algorithm to compute the Whittle index for the practical deployment of the policy. Although this paper focuses on UoI scheduling, our results apply to a general class of RMABs for which the UoI scheduling problem is a special case. Specifically, this paper's Whittle index policy is valid for any RMAB in which the bandits are binary Markov processes and the penalty is a concave function of the belief state of the Markov process. Numerical results demonstrate the excellent performance of the Whittle index policy for this class of RMABs. 

\end{abstract}

\begin{IEEEkeywords}
Uncertainty of information, RMAB, information freshness, AoI, scheduling.
\end{IEEEkeywords}

\section{Introduction}
\subsection{Information Freshness}
Fresh information is important in many modern information and control systems, particularly those used to support remote monitoring, industrial automation, and IoT applications. The concept of age of information (AoI) was first proposed as a metric of information freshness in 2011-12 \cite{AoI_originalpaper,AoI_2011}, and it has since attracted increasing attention. 
AoI measures the time elapsed since the generation of the latest packet delivered to the receiver. Many efforts have been devoted to designing minimum-AoI communication systems \cite{AoIdesign2017,Kadota2018TON,AoIdesign_henry2020,AoIdesign_sunyin2019,AoI2016Costa}. These investigations showed that using AoI as the performance metric would result in different system designs than using the traditional metrics of throughput and delay.

AoI assumes that the quality of information decreases with time in a way that is independent of the value of the last observation. In practice, given different initial states of a system, information quality may evolve with time in different ways. Consequently, how fast the last observation becomes outdated depends on the observed value. 
For example, consider a remote monitoring system with a discrete-time Markov process being observed. Assume that the Markov process has two states, say $0$ and $1$, and the state transition probabilities are $P[0|1]=0.3$ and $P[1|0]=1$. The transition probabilities are known to the monitor. If the monitor observes at time $t$ that the state of the process is $0$, then there is no need to schedule a new observation at time $t+1$ because the state of time $t+1$ must be $1$. In contrast, if the monitor observes at time $t$ that the state of the process is $1$, a new observation at time $t+1$ will help to reduce the uncertainty of the new state.
From the view of information theory, Shannon's metric for measuring uncertainty, i.e., entropy \cite{Cover_IT}, is the most fundamental way to measure how much we do not know about the latest source data in the absence of new information. The larger uncertainty of an old observation, the less ``useful information" it contains, hence the more urgent we need a new observation.
Motivated by this, we propose using the uncertainty of information (UoI) as a metric of information freshness and explore the optimal scheduling policy to minimize UoI.

There have been considerable efforts on using AoI as the performance metric in the design of scheduling strategies. Some studies also introduced a variety of non-decreasing functions of AoI as an extension to the AoI metric \cite{Symm_AoI,AoIcost_nonlinear2019,AoIcost_nonlinear_MIT,AoIaware18,MutualInform_sunyin}. For example, \cite{Symm_AoI} investigated scheduling for the minimum-cost of age in multi-source systems, where all sources have the same cost function that is non-decreasing with AoI.  
Perhaps the most relevant work to this paper, among the studies of AoI-based scheduling, is \cite{AoIcost_nonlinear_MIT}, where the authors proposed a Whittle index approach to minimizing the time-average of general non-decreasing functions of AoI. Compared with \cite{Symm_AoI}, the work in \cite{AoIcost_nonlinear_MIT} is more general because it deals with a model of diverse cost functions (i.e., each source may have different cost functions).  The relevance of these studies to our paper comes from the fact that UoI is a nonlinear function of AoI.
However, as will be elaborated in later sections, except for the special case where the system satisfies a certain symmetric property, UoI in general can be a decreasing or even an oscillating function of AoI, depending on the system dynamics. 
This is also a key difference between this paper and another relevant work. In fact, we are not the first to put forth an information-theoretic metric to the scheduling problems in information update systems. In \cite{LeiYing2020RMAB}, the authors studied a problem of monitoring multiple binary Markov chains over wireless channels with the aim of minimizing the information entropy. The system model and metric studied in \cite{LeiYing2020RMAB} are similar to this paper, but our problem is different in two ways: first, the objective function in \cite{LeiYing2020RMAB} is the discounted total entropy, while we aims to minimize the long-term average UoI. More importantly, the transition probabilities of each binary Markov chain in \cite{LeiYing2020RMAB} are assumed to be $P[0|1]=P[1|0]<0.5$, making the entropy an increasing function of AoI, as in \cite{AoIcost_nonlinear_MIT}. This paper, however, considers general Markov transition probabilities that $P[1|0]\le P[0|1]\in [0,1]$. As a result, UoI can be decreasing or even oscillating with AoI. 
Therefore, our problems fall outside the scope of treatments in the existing work. The challenges brought about by the generalization on the Markov transition probabilities are two folds. First, if $P[0|1]\neq P[1|0]$, then UoI depends not only on AoI, but also on the state of the last observation. Second, a Markov chain with $P[0|1]+ P[1|0]>1$ has different characteristics in the evolution of multi-step transition probability than the case of $P[0|1]+ P[1|0]<1$, making the analytical treatment much more challenging.

\subsection{Restless Multi-armed Bandit}

In this work, we put forth a restless multi-armed bandit (RMAB) formulation for the problem of minimum sum-UoI scheduling. RMAB is a generalization of the classical multi-armed bandit (MAB) problem  \cite{gittins_RMAB_book}. MAB concerns the activation of $n$ bandit processes. At each time step, only one process is to be activated. If a process is activated at time $t$, then a reward is generated and the state of the process changes according to a Markov rule specific to the process. The processes that are not activated receive no reward, and their states do not change. 
Whittle generalized MAB to RMAB in two ways \cite{RMAB_Whittle1988}. First, at each time step, $m$ out of the $n$ bandit processes can be activated ($1\le m<n$). Second, the unactivated $n-m$  processes may also have rewards and their states may change (hence, the term ``restless'' in RMAB).

It is well-known that the Gittins index policy is optimal for MAB \cite{gittinsJones}. For an index policy, an index is assigned to each bandit process at each time step, and the bandit process with the largest index is activated.
The index policy making use of the Gittins index is suboptimal for RMAB. Indeed, it has been shown that RMAB  is PSPACE-hard \cite{RMAB_PSPACEhard}, and RMAB is much harder than MAB as far as an optimal policy is concerned.

Whittle introduced a Lagrange multiplier to relax the RMAB problem and developed an index policy using a heuristic index, commonly referred to as the Whittle index \cite{RMAB_Whittle1988}. Intuitively, Whittle index evaluates how rewarding it is to activate a bandit process in its current state. The $m$ bandit processes with the largest $m$ indices are activated. The Whittle index policy has been widely used in RMAB problems and has exhibited near-optimal performance in extensive applications \cite{Whittle_app2003,Whittle_app2006,Whittle_app2018,Shiling2020}. However, the Whittle index policy is not naturally applicable to all RMAB problems---only an ``indexable'' RMAB has a well-defined Whittle index. In general, the indexability of an RMAB can be challenging to establish, and the Whittle index can be complicated to compute. 

Many research efforts on RMABs focus on the establishment of indexability and the computation of Whittle index \cite{nino_2001,nino2007dynamic,Index_compute2020,Liu2010,villar_2016}. For example,
 \cite{nino_2001} and \cite{nino2007dynamic} studied the sufficient conditions for the indexability of RMABs based on the achievable region method. However, the results are not applicable to our problem because the RMABs studied only cover bandits with finite states and linear rewards. 
A work that is relevant to ours is \cite{Liu2010}. The authors in \cite{Liu2010} studied a class of RMABs motivated by dynamic multichannel access. In \cite{Liu2010}, each bandit process is modeled as a partially observable Markov decision process (POMDP), and it receives a reward only when it is activated. The reward is a linear function of the belief state. The authors established the indexability and gave a closed-form Whittle index for this class of RMABs.  These results, however, do not apply to our UoI-scheduling problem because (i) UoI is a concave function of the belief state; (ii) UoI is incurred as a penalty (i.e,  negative reward) even when a bandit is not activated in our problem.

\subsection{Main Results}
This paper considers a system in which a central monitor observes $M$ binary Markov processes through $m$ reliable communication channels. At each time step, only the states of $m$ Markov processes can be communicated to the central monitor. Scheduling refers to the selection of the $m$ Markov processes at each time step. We study the scheduling policy that minimizes the time-averaged sum-UoI of the Markov processes by formulating each bandit process as a POMDP under the RMAB framework. The penalty (i.e., UoI) of each bandit process is a concave function of the belief state. The study of the decoupled single-bandit problem is an essential step toward establishing the RMAB’s indexability. Toward that end, we first analyze the properties of the optimal policy for the POMDP associated with a single bandit process. We show that the optimal policy has a threshold structure that allows us to establish the indexability of our UoI-scheduling RMAB problem. We further present an iterative algorithm to compute the Whittle index. In the special case where the Markov process has symmetric transition probabilities, UoI reduces to an increasing function of AoI and a closed-form expression of Whittle index can be derived.

We emphasize that the results in this paper apply not only to the UoI-scheduling problem, but also to a large class of RMABs (we call them C-type RMABs). In essence, this work covers a general RMAB wherein each bandit process is a binary Markov process with transition probabilities $p$ and $q$, and the penalty is a concave function (not limited to UoI) of the belief state \cite{Mauricio_POMDP2010}.
As we will see, a bandit process with $p+q<1$ and a bandit process with $p+q>1$ have different characteristics, but both of them are indexable.

In summary, the contributions of this paper are as follows:
\begin{itemize}
	\item We propose using UoI as a metric for information freshness, and formulate the minimum sum-UoI scheduling problem as an RMAB.
	\item We prove that a general class of RMABs---to which the UoI-scheduling problem belongs---is indexable as long as the penalties are concave functions of the bandits' belief states.
	\item We present an iterative algorithm to compute the Whittle index of the RMABs and derive the closed-form expression of Whittle index in a special case.
\end{itemize}

The rest of this paper is organized as follows. Section \ref{sec:2} presents the system model and the formulation of the UoI-scheduling problem. Section \ref{sec:3} introduces the RMAB formulation and the basic concepts of the Whittle index policy. Section \ref{sec:singlebandit} studies the single-bandit problem and develops useful properties for later development. Section \ref{sec:index} establishes the indexability of our RMAB. Section \ref{sec: alg} develops an iterative algorithm to compute the Whittle index. Section \ref{sec:discussion} studies a special case and discusses possible directions for future work. Section \ref{sec:simul} presents simulation results that demonstrate the excellent performance of our method. Finally, Section \ref{sec:con} concludes this paper. 

\subsection{Notations}
$\mathbb{N}$ denotes the set of natural numbers, and $\mathbb{N}^+$ denotes the set of positive integers. For a positive integer $M$, $[M]$ denotes the the collection of integers between $1$ and $M$, i.e., $[M]\triangleq \{1,\cdots,M\}$. $P[\cdot |\cdot]$ denotes the conditional probability.

\begin{figure}[t]
	\centering
	\includegraphics[width=3.2in]{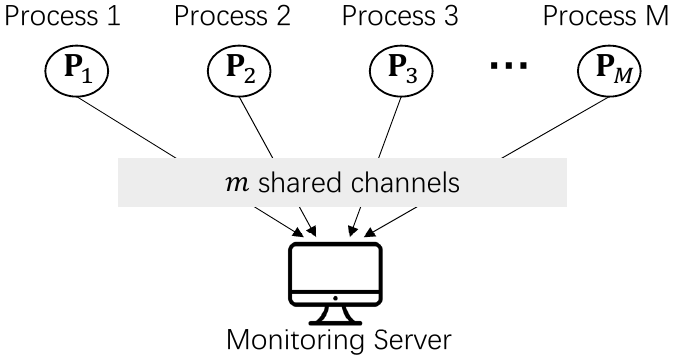}
	\caption{System Model: $M$ remote processes being monitored by a monitoring server over $m$ shared channels.}
	\label{fig:sys_model}	
\end{figure}
\section{Problem Statement}  \label{sec:2}
\subsection{System Model}
Consider a system with $M$ remote processes being observed by a centralized monitoring server, as shown in Fig. \ref{fig:sys_model}. The states of the remote processes are delivered to the monitoring server over a set of $m$ shared channels, $1\le m <M$. We assume all the $m$ channels are reliable (specifically, the probability of successful transmission is 1) and operate in a time-slotted manner. At the beginning of each time slot, $m$ processes are selected to send their current states to the monitoring server. The monitoring server receives the state information at the end of the slot.

The remote processes are independent discrete-time binary Markov processes with states at time $t$ denoted by $S_i(t)\in \{0,1\}, i\in [M]$. The states of different processes evolve in time at different rates according to the state-transition probabilities $P[{S_i}(t + 1)|{S_i}(t)], i \in [M]$. A discrete time step of the Markov processes corresponds to one time slot of the communication channels, and $S_i(t)$ is the state at the beginning of time slot $t$. Thus, at the end of time slot $t$, the Markov processes would have evolved to states $S_i(t+1)\in \{0,1\}, i\in [M]$ (the end of time slot $t$ corresponds to the beginning of time slot $t+1$ in continuous time).

The one-step transition matrix of each remote Markov process is known to the monitoring server. In particular, assume the transition matrix of process $i \in [M]$ is given by
\begin{align*}
\mathbf{P}_i\triangleq \begin{bmatrix}
{P [ 0| 0]}&{P [ 0| 1]}\\
{P [ 1| 0]}&{P [1| 1]}
\end{bmatrix} = \begin{bmatrix}
{1 - {p_i}}&{{q_i}}\\
{{p_i}}&{1 - {q_i}}
\end{bmatrix}.
\end{align*}
Without loss of generality, we assume that $0\le p_i \le q_i \le1$ for all $i\in [M]$. We can easily obtain the $n$-step transition matrix of this Markov process:
\begin{align} \label{eq: nstep_prob}
\mathbf{P}_i^n = \begin{bmatrix}
1 - p_i^{(n)}&q_i^{(n)}\\
p_i^{(n)}&1 - q_i^{(n)}
\end{bmatrix},  
\end{align}
where
\begin{align*}
p_i^{(n)} \triangleq \frac{{{p_i} - {p_i}{{(1 - {p_i} - {q_i})}^n}}}{{{p_i} + {q_i}}}, \; q_i^{(n)}\triangleq \frac{{{q_i} - {q_i}{{(1 - {p_i} - {q_i})}^n}}}{{{p_i} + {q_i}}}.
\end{align*}
Note that $p^{(n)}_i$ denotes the n-step transition probability from state 0 to state 1. It should not be confused with $p^n_i$.

\begin{figure*}[!t]
	\centering
	\begin{subfigure}[b]{0.32\linewidth}
		\includegraphics[width=\linewidth]{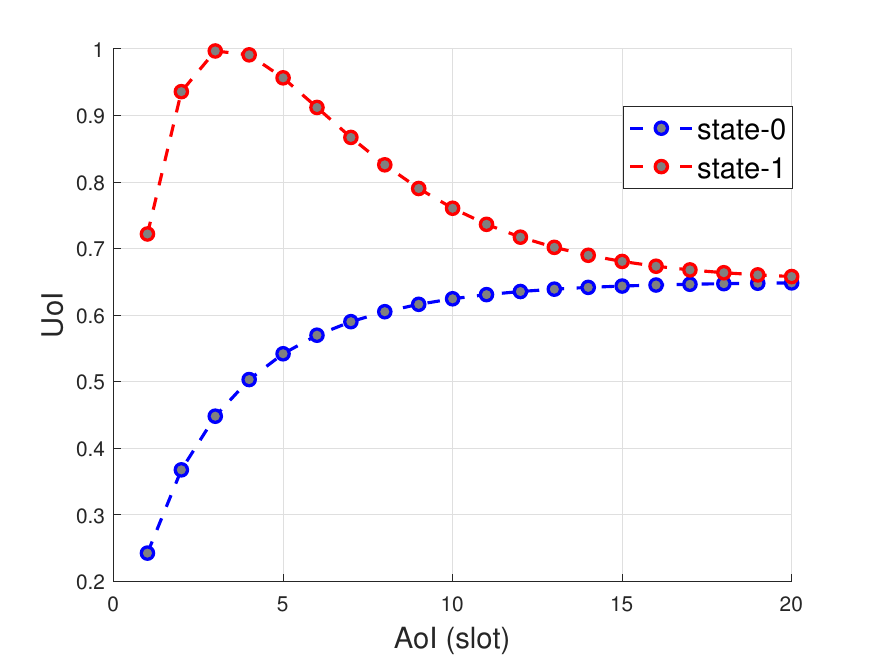}
		\caption{$p_1=0.04,q_1=0.2$.}
		\label{fig:examp1}
	\end{subfigure}
	\begin{subfigure}[b]{0.32\linewidth}
		\includegraphics[width=\linewidth]{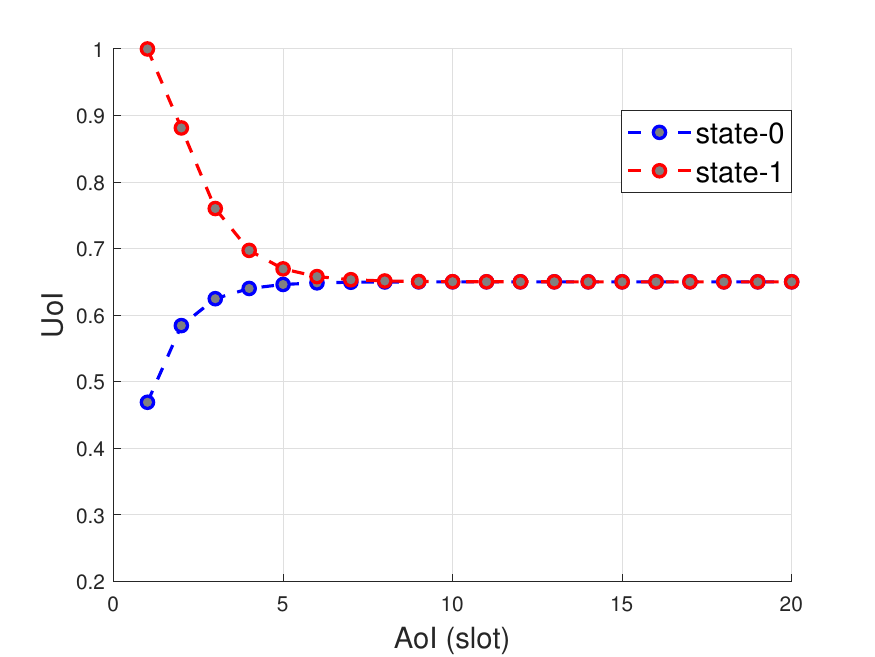}
		\caption{$p_2=0.1,q_2=0.5$.}
		\label{fig:examp2}
	\end{subfigure}
	\begin{subfigure}[b]{0.32\linewidth}
		\includegraphics[width=\linewidth]{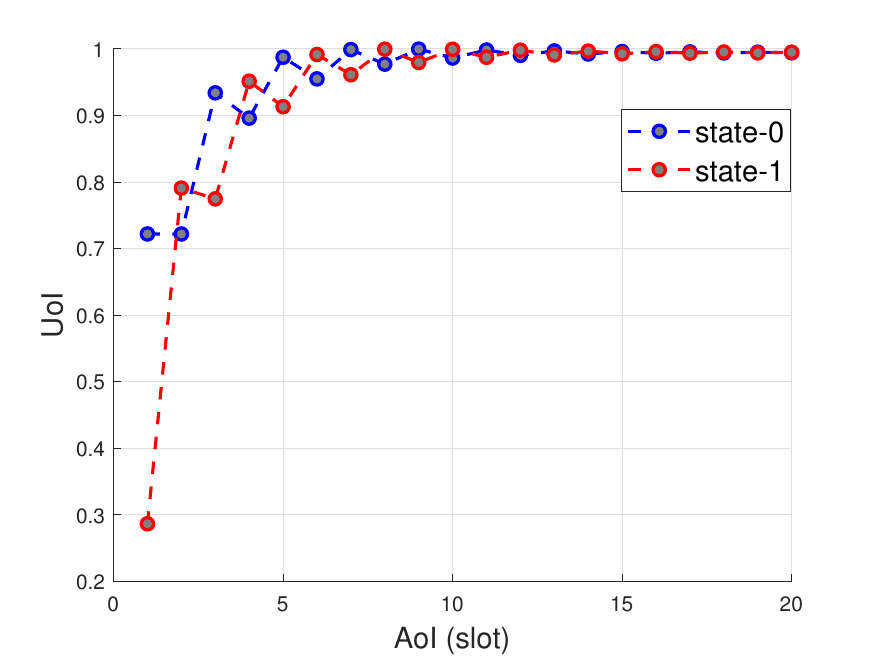}
		\caption{$p_3=0.8,q_3=0.95$.}
		\label{fig:examp3}
	\end{subfigure}	
	\caption{UoI as a function of AoI, for (a) remote process 1 with one-step transition probabilities $p_1=0.04,q_1=0.2$; (b) remote process 2 with one-step transition probabilities $p_2=0.1,\ q_2=0.5$; (c) remote process 3 with one-step transition probabilities $p_3=0.8,\ q_3=0.95$. In the legend, state-0 means that the last observation is 0; state-1 means that the last observation is 1.}
	\label{fig:example}
\end{figure*}

\subsection{Uncertainty of Information}
In this paper, we define information staleness in terms of uncertainty. In information theory, the uncertainty of information is measured by Shannon’s entropy. Specifically, at the end of slot $t$, if the latest observation of process $i$ at the monitoring server is $S_i(t')$ ($t'\le t$), then the UoI of process $i$ at the end of slot $t$, denoted by $U_i[t]$, is the entropy of $S_i(t+1)$ conditioned on the given $S_i(t')$: 
\begin{align} \label{eq:UoI-1}
{U_i}[t] = - \sum\limits_{{S_i}(t+1)} {P[{S_i}(t+1)|{S_i}(t')]{{\log }_2}} P[{S_i}(t+1)|{S_i}(t')],
\end{align}
where $P[{S_i}(t+1)|{S_i}(t')]$ is the  transition probability of the state from time $t'$ to $t+1$. Suppose that $t'=t-n+1$ , where $n\in \{1,2,...,t\}$ is the time elapsed since the generation of $S_i(t')$. Then $P[{S_i}(t+1)|{S_i}(t')]$ is the $n$-step transition probability given by \eqref{eq: nstep_prob}. 
Define the entropy function
\begin{align}
H(p) \triangleq -p \log_2(p) -(1-p)\log_2(1-p).
\end{align}
We thus rewrite \eqref{eq:UoI-1} as follows:
\begin{align}  \label{eq:UoI-2}
U_i[t]= \begin{cases}
H\big(p_i^{(n)}\big), \quad \text{if } S_i(t')=0  \\
H\big(q_i^{(n)}\big), \quad \text{if } S_i(t')=1.
\end{cases}
\end{align}
Note that $H(p)$ is concave w.r.t. $p\in [0,1]$ and it reaches its maximum at $p=0.5$. According to \eqref{eq:UoI-2} and \eqref{eq: nstep_prob}, given $S_i(t')$, UoI may not be a monotonic function of the age (i.e., $n$). Fig. \ref{fig:example} shows some examples with different transition probabilities. In particular, if $p_i + q_i<1$, then $H\big(p_i^{(n)}\big)$ is increasing w.r.t. $n$, while $H\big(q_i^{(n)}\big)$ may be decreasing w.r.t. $n$. If $p_i + q_i>1$, then $H\big(p_i^{(n)}\big)$ and $H\big(q_i^{(n)}\big)$ oscillate as $n$ increases. For the special case of $p_i + q_i = 1$, $p^{(n)}_i=p_i$ and $q^{(n)}_i=q_i$ for all $n$; hence $U_i[t]=H(p),\forall t$. That is, the UoI is a constant that can not be changed by any scheduling policy. We thus assume $p_i+q_i\neq 1$ for all $i\in [M]$ in this study.  For a similar reason, we also assume that $p_i+q_i\neq 0$ and $p_i+q_i\neq 2$ for all $i\in [M]$.

In each time slot, only the states of $m$ remote processes can be delivered over the shared channels to the monitor. Hence, there is a trade-off among the remote processes' UoIs at the monitoring server. This paper studies the scheduling of the updates of the remote processes to minimize the average sum-UoI over the infinite horizon:
\begin{align} \label{Pro:UoImin}
\min {\rm{  }}\mathop {\lim }\limits_{T \to \infty } \frac{1}{T}\sum\limits_{t = 1}^T {\sum\limits_{i = 1}^M {{U_i}[t]} }. 
\end{align}

Since UoI may not be an increasing function of AoI, UoI-based scheduling is different from the AoI-based scheduling studies in many prior works \cite{Symm_AoI,AoIcost_nonlinear2019,AoIcost_nonlinear_MIT,AoIaware18,MutualInform_sunyin}. How to schedule the updates of remote processes to minimize the average sum-UoI is a new and unexplored problem.

\section{Restless Multi-armed Bandit and Index Policy}  \label{sec:3}
This section formulates the minimum sum-UoI scheduling problem as an RMAB. In addition, we introduce the basic concepts of the Whittle index policy, a widely used algorithm for RMAB problems.

\subsection{Restless Multi-armed Bandit Formulation}
\begin{figure}[!t]
	\centering
	\begin{subfigure}[b]{0.48\linewidth}
		\includegraphics[width=\linewidth]{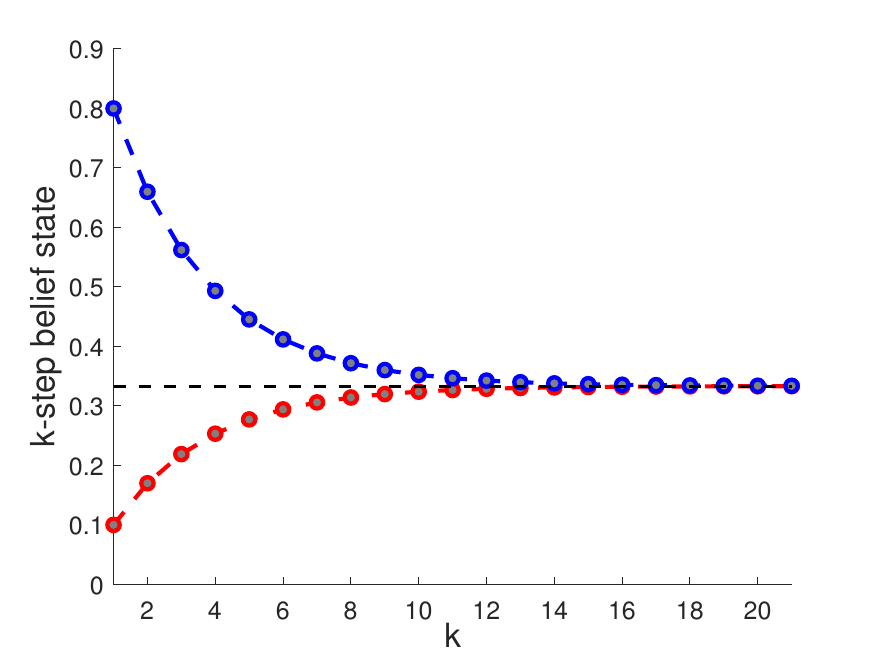}
		\caption{$p_i+q_i<1$.}
		\label{fig:belief_update1}
	\end{subfigure}
	\begin{subfigure}[b]{0.48\linewidth}
		\includegraphics[width=\linewidth]{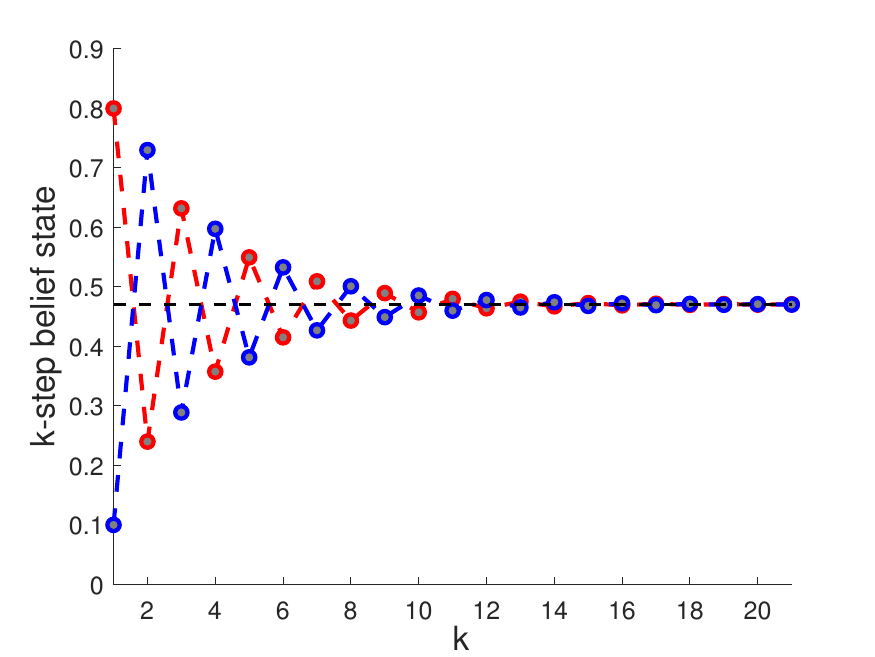}
		\caption{$p_i+q_i>1$.}
		\label{fig:belief_update2}
	\end{subfigure}
	\caption{$k$-step belief state evolution under the passive action.}
	\label{fig:belief_update}
\end{figure}

We formulate the problem of scheduling the updates of the $M$ remote processes as an RMAB with $M$ bandit processes, each corresponding to one remote process. Let $u_i(t)\in \{0,1\}$ denote the action applied to bandit $i$ in slot $t$, where $u_i(t)=1$ (active action) means process $i$ is selected to transmit in slot $t$ and $u_i(t)=0$ (passive action) otherwise. 
We use the ``belief state'' $\omega_i(t)$ \cite{Mauricio_POMDP2010} to represent the state of bandit $i$ at the beginning of slot $t$. Specifically, for bandit $i$, $\omega_i(t)\in [0,1]$ is the probability that $S_i(t)=1$. Given the action $u_i(t)$ and the observation $S_i(t)$, the belief state at the beginning of slot $t+1$ is given recursively as follows:
\begin{align}
{\omega _i}(t + 1) =  \begin{cases}
p_i ,\ &\text{if } u_i(t)=1 \text{ and } S_i(t)=0\\
1-q_i ,\ &\text{if } u_i(t)=1 \text{ and } S_i(t)=1 \\
\tau(\omega_i(t)),\  &\text{if } u_i(t)=0
\end{cases} 
\end{align}
where $i\in [M]$ and 
\begin{align}
\tau(\omega_i(t))\triangleq p_i + \omega_i(t)(1-p_i-q_i).
\end{align}
Operator $\tau(\cdot)$ is the one-step belief state evolution under the passive action. We further define the $k$-step belief state evolution ($k>1$):
\begin{align}
\tau^k(\omega_i) \triangleq \tau\big(\tau^{k-1}(\omega_i)\big).
\end{align}
As shown in Fig. \ref{fig:belief_update}, the $k$-step belief state evolutions for a bandit with $p_i+q_i<1$ and for a bandit with $p_i+q_i>1$ are quite different. Some important properties of the belief state evolution under consecutive passive actions are stated in Lemma 1 below.

\begin{lemma} \label{Lem:1}
	Let $p_i,q_i \in [0,1]$ denote the transition probabilities of a bandit $i$. For any $\omega_i \in [0,1]$ and $k\in \mathbb{N}^+$, we have
	\begin{align}  \label{eq:lemma1-1}
	\mathop {\lim }\limits_{k \to \infty } {\tau ^k}({\omega _i}) &= \omega _i^* \triangleq  \frac{{{p_i}}}{{{p_i} + {q_i}}},  \\  \label{eq:lemma1-2}
	\frac{{{\tau ^k}({\omega _i}) - \omega _i^*}}{{{\omega _i} - \omega _i^*}} &= {(1 - {p_i} - {q_i})^k},
	\end{align}
	where $\omega _i^*$ is the equilibrium belief state in the lack of observations throughout the time horizon. Furthermore, the convergence of $\tau^k(\omega_i)$ to $\omega_i^*$ has the following properties:
	\begin{itemize}
		\item[(1)] If $p_i+q_i<1$ (monotonic bandit), for any $\omega_i \in [0,1]$, $\tau^k(\omega_i)$ monotonically converges to $\omega_i^*$ as $k \to \infty$.
		\item[(2)] If $p_i+q_i>1$ (oscillating bandit), for any $\omega_i \in [0,1]$, $\tau^k(\omega_i)$ oscillatorily  converges to $\omega_i^*$ as $k \to \infty$. Specifically, $\tau^{2k}(\omega_i)$ and $\tau^{2k+1}(\omega_i)$ converges to $\omega_i^*$ from opposite directions.
	\end{itemize}
\end{lemma}
\begin{IEEEproof}
	For any $k\in \mathbb{N}^+$, we have
	\begin{align*}
	{\tau ^k}({\omega _i}) &= \sum\limits_{n = 0}^{k-1} {{p_i}{{(1 - {p_i} - {q_i})}^n}}  + {\omega _i}{(1 - {p_i} - {q_i})^k} \\
	&= \frac{{{p_i} - {p_i}{{(1 - {p_i} - {q_i})}^k}}}{{{p_i} + {q_i}}} + {\omega _i}{(1 - {p_i} - {q_i})^k}.
	\end{align*}
	Since $0< |1-p_i-q_i|<1$, \eqref{eq:lemma1-1} is immediate from the above equation by letting $k \to \infty$. For finite $k$, the above equation and \eqref{eq:lemma1-1} give \eqref{eq:lemma1-2}.	
\end{IEEEproof}

With the RMAB model, we can rewrite the problem \eqref{Pro:UoImin} as follows:
\begin{align} \label{Pro:P1}
{\rm{P1:  }}\quad &\min\limits_{\{ {u_i}(t)\} } \ \lim \limits_{T \to \infty } \frac{1}{T}\sum\limits_{t = 1}^T {\sum\limits_{i = 1}^M {H\left( {{\omega _i}(t)} \right)} }  \\  \label{Pro:P1st}
&\ \ s.t.\quad \ \sum\limits_{i = 1}^M {{u_i}(t) = m, \ \forall t}  \\
& \qquad \quad \  u_i(t) \in \{0,1\}, \ \forall i,t. 
\end{align}

RMAB in general is PSPACE-hard and we can not expect to find the optimal solution easily \cite{RMAB_PSPACEhard}. To solve RMAB effectively, Whittle introduced a method to relax RMAB \cite{RMAB_Whittle1988}. The relaxed problem can then be decoupled to $M$ single-bandit problems to be examined separately to compute a ``Whittle index'' for each of them. In our case, the relaxed scheduling problem consists in selecting the $m$ bandits with the largest $m$ Whittle indices for states update in each time slot.  This is referred to as an ``index policy''. The validity of the index policy depends on the underlying problem being indexable. The indexability property will be elaborated in Section III.B, and we establish in Section V that our UoI-scheduling problem is indexable.

Constraint \eqref{Pro:P1st} means that exactly $m$ out of the $M$ bandits can take active action in each slot. Using the Whittle approach, it is relaxed in the following way:
\begin{align} \label{eq:relaxst}
\mathop {\lim }\limits_{T \to \infty } \frac{1}{T}\sum\limits_{t = 1}^T {\sum\limits_{i = 1}^M {{u_i}(t)} } = m.
\end{align}
Note that constraint \eqref{eq:relaxst} only requires the number of bandits taking the active action in each slot, averaged over time, to be $m$. Replacing \eqref{Pro:P1st} with \eqref{eq:relaxst}, we get a relaxation of problem P1, which can be further transformed by a Lagrange multiplier $\lambda$:
\begin{align*}
{\rm{P2:  }}\mathop {\min }\limits_{\{ {u_i}(t) \} } {\rm{  }}\mathop {\lim }\limits_{T \to \infty } \frac{1}{T}\sum\limits_{t = 1}^T {\left[ {\sum\limits_{i = 1}^M {H\left( {{\omega _i}(t)} \right)} {\rm{ + }}\lambda \sum\limits_{i = 1}^M {{u_i}(t)} } \right]}  - m\lambda, 
\end{align*}
where $u_i(t)\in \{0,1\}$. By interchanging the summation over $t$ and the summation over $i$, we can decouple the above problem to $M$ subproblems as follows:
\begin{align} \label{Pro:singlebandit}
J_i: = \mathop {\min }\limits_{\{ {u_i}(t)\} } {\rm{  }}\mathop {\lim }\limits_{T \to \infty } \frac{1}{T}\sum\limits_{t = 1}^T {\left[ {H\left( {{\omega _i}(t)} \right) + \lambda {u_i}(t)} \right]} ,{\rm{  }}i \in [M]
\end{align}
Each subproblem ${J}_i$  is the objective of a single bandit associated with remote process $i$. The multiplier $\lambda$ is non-negative and can be interpreted as a service charge for taking the active action, and it is incurred each time the active action is applied to the bandit. 

Solving the relaxed problem P2 does not provide the exact optimal solution for P1. The advantage of considering P2, however, is that we can decompose the original $M$-dimensional problem into $M$ independent 1-dimensional problems. It turns out that the Whittle index policy developed based on the decoupled single-bandit problems can achieve near-optimal performance for many RMAB problems \cite{Whittle_app2003,Whittle_app2006,Whittle_app2018,Shiling2020}.

\subsection{Whittle Index Policy and Indexability}
Whittle index policy computes an index for each belief state $\omega_i$ of a bandit $i$. At the beginning of each slot, the policy activates the $m$ bandits with the largest $m$ indices. The Whittle index of a bandit $i$ only depends on the parameters $p_i$ and $q_i$ associated with the bandit, and it is obtained by considering the single-bandit problem of \eqref{Pro:singlebandit}. 

For any fixed service charge $\lambda$, the optimal policy of problem $J_i$ partitions the belief state space into a passive set $E_0^i$ and an active set $E_1^i$, where, respectively, the optimal action is $u_i=0$ (passive) and $u_i=1$ (active). For definiteness, if it is equally optimal to take the two actions in a belief state $\omega_i$, we will let $\omega_i \in E_0^i$. That is, the bandit will take passive action in this belief state. Note that $E_0^i$ and $E_1^i$ would vary with $\lambda$, we thus use $E_0^i(\lambda)$ and $E_1^i(\lambda)$ to denote the passive and active sets with service charge $\lambda$. 
We now introduce the definition of indexability.

\begin{definition}[Indexability \cite{gittins_RMAB_book}] 
	A bandit $i$ is indexable if the passive set $E_0^i(\lambda)$ with service charge $\lambda$ monotonically expands from the empty set to the whole belief state space as $\lambda$  increases from 0 to $+\infty$. Specifically, $E_0^i(\lambda_1)\subseteq E_0^i(\lambda_2)$ if $\lambda_1<\lambda_2$. An RMAB is indexable if all bandits are indexable. 
\end{definition}

If the RMAB problem P1 is indexable, a Whittle index policy for this problem can be developed by determining the Whittle index defined as follows.
\begin{definition}[Whittle index \cite{gittins_RMAB_book}]
	If bandit $i$ is indexable, the Whittle index of this bandit in belief state $\omega_i$, denoted by $W(\omega_i)$, is the service charge $\lambda$ making the two actions equally rewarding for bandit $i$ in belief state $\omega_i$. Equivalently, $W(\omega_i)$ is the infimum $\lambda$ such that it is optimal to take the passive action in belief state $\omega_i$. That is
	\begin{align*}
	W({\omega _i}) = \mathop {\inf }\limits_\lambda  \left\{ {\lambda :{\omega _i} \in E_0^i(\lambda )} \right\}.
	\end{align*}
\end{definition}

Intuitively, the Whittle index evaluates how rewarding it is to activate a bandit in a particular belief state. Hence the $m$ bandits with the largest $m$ indices should be selected to take the active action in that slot. By definition, the Whittle index can be computed by solving the associated single-bandit problem. Therefore, in the following sections, we will focus on a general single-bandit problem. We first present properties of the optimal policy for the single-bandit problem, based on which we establish the indexability of our problem. After that, we develop an algorithm to compute the Whittle index.

\section{The Single-bandit Problem} \label{sec:singlebandit}
This section studies the single-bandit problem of \eqref{Pro:singlebandit}. For simplicity, we drop the bandit index from all notations in this section. For example, $p_i$ and $q_i$ will be simply expressed as $p$ and $q$.

\subsection{Optimality Equations for the Single-bandit Problem}
We reformulate the single-bandit problem as an MDP with continuous state space $[0,1]$ and finite action space $\{0,1\}$. The belief state space of the single-bandit problem is a countably infinite set, $\Omega  \triangleq  \{ {p^{(n)}},1 - {q^{(n)}}:n \in \mathbb{N}^+\} $. Given belief state $\omega(t)$ and action $u(t)$ in slot $t$, the state-transition probability that $\omega(t+1)=\omega'$ is given by
\begin{align} \label{eq:bandit-transit}
&P[\omega(t+1)=\omega'|\omega(t),u(t)] \nonumber\\ 
=&  \begin{cases}
\omega(t), &\text{if } u(t)=1, \omega'=1-q\\
1-\omega(t), &\text{if } u(t)=1,  \omega'=p \\
1,  &\text{if } u(t)=0, \omega'=\tau(\omega(t)) \\
0,  &\text{otherwise.}
\end{cases} 
\end{align}
The penalty function for the state-action pair $(\omega(t),u(t))$ is $H(\omega(t))+\lambda u(t)$. The objective is to minimize the average penalty over the infinite horizon, as given by \eqref{Pro:singlebandit}. 

The above single-bandit problem is a multichain MDP. To see this, let us first define unichain and multichain policies, and unichain and multichain MDPs, as follows \cite{puterman1994markov}:
\begin{definition}[Unichain and Multichain Policies]
	A deterministic stationary policy is called a unichain policy if the Markov process corresponding to the policy is unichain. That is, the Markov process consists of a single recurrent class plus some transient states. If a deterministic stationary policy is not unichain, we call it a multichain policy.
\end{definition}

Two states that communicate are said to be in the same class. Furthermore, two states are said to be in the same chain if they belong to, or can transit to, the same class. A multichain policy results in multiple chains.
\begin{definition}[Unichain and Multichain MDPs]
	An MDP is unichain if all deterministic stationary policies are unichain policies; if there exists at least one deterministic stationary policy that is multichain, then the MDP is multichain.	
\end{definition}

The following example is enough to show that the single-bandit MDP is multichain.

\begin{example}
	Consider a policy that takes the passive action in all belief states $\omega \in [p^{(2)},1-q^{(2)}]$ and the active action otherwise. Note that the equilibrium belief state $\omega^* \in [p^{(2)},1-q^{(2)}]$. Under this policy, the belief states are divided into two chains: (1) belief states belonging to $[p^{(2)},1-q^{(2)}]$ will evolve to and stay in $\omega^*$; (2) belief states not belonging to $[p^{(2)},1-q^{(2)}]$ will stay within a recurrent class consisting of $p$ and $1-q$.
\end{example}

Since the single-bandit problem is a multichain MDP, the optimal policy is determined by a set of optimality equations as follows \cite{puterman1994markov}:
\begin{align} \label{eq:firstopt_eq}
&\mathop {\min }\limits_{u \in \{ 0,1\} } \big\{ \omega g(1 - q) + (1 - \omega )g(p),g\left( {\tau (\omega )} \right)\big\}  = g(\omega ), \\
&H(\omega ) +  \min\limits_{u \in {A_\omega }} \{ \lambda  + \omega V(1 - q) + (1 - \omega )V(p),V\left( {\tau (\omega )} \right)\}\nonumber \\ \label{eq:bellmaneq}
&=V(\omega ) + g(\omega ), 
\end{align}
where $g(\omega)$ is the average penalty with initial state $\omega$, $V(\omega)$ is the value function related to the asymptotic total difference between the cumulative penalty and the stationary penalty. In \eqref{eq:bellmaneq}, $A_{\omega}$ is a subset of the action space that satisfies \eqref{eq:firstopt_eq}, i.e.,
\begin{align*}
A_\omega &= \big\{u\in \{0,1\}: \\ 
&u[\omega g(1 - q) + (1 - \omega )g(p)]+(1-u)g(\tau(\omega))=g(\omega) \big\}.
\end{align*}
We refer to \eqref{eq:firstopt_eq} as the first optimality equation, and \eqref{eq:bellmaneq} as the second optimality equation or the Bellman equation. To determine the optimal policy of the single-bandit problem, we need to find $V(\omega)$ and $g(\omega)$ that satisfy \eqref{eq:firstopt_eq}-\eqref{eq:bellmaneq} for all $\omega \in \Omega$. In the following, we extend the belief state space to the continuous region $[0,1]$ to ease analysis. After doing so, we aim at finding $V(\omega)$ and $g(\omega)$ that satisfy \eqref{eq:firstopt_eq}-\eqref{eq:bellmaneq} for all $\omega \in [0,1]$. Since $\Omega \subset [0,1]$, as far as the optimal policy is concerned, the value function for the extended belief state space leads to the optimal policy for the single-bandit problem.

Starting from any $\omega(0)\neq \omega^*$, \eqref{eq:bandit-transit} and Lemma 1 indicate that $\omega(t)$ will approach, under the passive action, the equilibrium belief state $\omega^*$ asymptotically, but can never reach  $\omega^*$ within a finite time. Hence it is impractical to implement a policy that takes different actions in $\omega^*$ and other belief states in $B_\epsilon(\omega^*)\triangleq \{ \omega :|\omega  - {\omega ^ * }| < \epsilon \} $ as $\epsilon\to0$. We thus make the following assumption to to exclude policies that take different actions in $\omega^*$ and its neighborhood:

\textit{If a policy takes active (passive) action in $\omega^*$, it must take active (passive) action in every $\omega  \in  B_\epsilon(\omega^*) $ for some $\epsilon>0$.} 

 With this assumption, a policy is not considered an admissible policy if, for every $\epsilon>0$, there exists a $\omega_\epsilon  \in  B_\epsilon(\omega^*) $ such that the policy takes different actions in $\omega^*$ and $\omega_\epsilon$.

For the single-bandit problem, all multichain policies are two-chain policies, with properties stated in Lemma 2 below:

\begin{lemma} \label{Lem:2}
	For the single-bandit problem, the Markov process corresponding to any  multichain policy consists of two chains. Such a multichain policy 
	\begin{itemize}
		\item [1.] takes the passive action in $\omega^*$; and
		\item [2.] takes the active action in $p^{(n)}$ and $1-q^{(m)}$ for some finite integers $n$ and $m$, respectively.
	\end{itemize} 
Furthermore, a multichain policy partitions the belief states into two chains, where
\begin{itemize}
	\item Chain 1 includes belief states $p$  and $1-q$ among other possible states in the same chain.
	\item Chain 2 includes the equilibrium belief state $\omega^*$ among other possible states in the same chain.
\end{itemize}
\end{lemma}
\begin{IEEEproof}
	See Appendix \hyperref[Apd:A]{A}.	
\end{IEEEproof}

\subsection{Properties of the Optimal Policy}
This part presents properties of the optimal policy for the single-bandit problem that play a fundamental role in establishing indexability. In general, a multichain MDP is more difficult to tackle than a unichain MDP. Fortunately, we find that the single-bandit problem can be optimized by a unichain policy. 
\begin{lemma} \label{Lem:3}
	The optimality of the single-bandit problem can be achieved by a unichain policy.
\end{lemma}
\begin{IEEEproof}
	To prove this lemma, it suffices to show the following: For any multichain policy, we can find a unichain policy that is not worse than the multichain policy. The complete proof is given in Appendix \hyperref[Apd:A]{A}.	
\end{IEEEproof}

With Lemma 3, we can restrict our attention to unichain policies. The average penalty of a unichain policy does not vary with the initial state, i.e., $g(\omega)=g$ is a unique constant. Hence a unichain policy always satisfies the first optimality equation. In this case, the first optimality equation is redundant and can be removed. We then focus on the Bellman equation. The following lemma states that the value function of the optimal unichain policy is concave.
\begin{lemma} \label{Lem:4}
	For $V(\omega)$ and $g(\omega)$ that satisfy the Bellman equation:
	\begin{align*}  
	&H(\omega ) +  \min \big\{ \lambda  + \omega V(1 - q) + (1 - \omega )V(p),V\left( {\tau (\omega )} \right)\big\} \\
	&=V(\omega ) + g(\omega ), 
	\end{align*}
	the optimal penalty $g(\omega)$ does not vary with $\omega$ (expressed as $g(\omega)=g$). In addition, $V(\omega)$ is a concave function of $\omega$.
\end{lemma}
\begin{IEEEproof}
	Since the single-bandit problem can be optimized by a unichain policy, the optimal penalty $g(\omega)$ does not vary with $\omega$. We next prove the concavity of $V(\omega)$ based on the convergence of relative value iteration.  
	Let $Z_0(\omega)=V_0(\omega)=0$. Fix a reference state $\omega_r$ (e.g., $\omega_r=p$). For $n\in \mathbb{N}$, define the following recursion:
	\begin{align} \label{eq:relatedVI1}
	&Z_{n + 1}(\omega ) = \mathcal{L}{V_n}(\omega ) \triangleq H(\omega ) + \nonumber \\
	& \min \{ \lambda  + \omega {V_n}(1 - q) + (1 - \omega ){V_n}(p),{V_n}(\tau (\omega ))\}, \\
	&g_{n+1} = Z_{n+1}(\omega_r),  \\ \label{eq:relatedVI3}
	&V_{n+1}(\omega)=\mathcal{L}V_n(\omega) - g_{n+1},
	\end{align}
	where $\mathcal{L}$ is the Bellman operator. It turns out that the sequences $\{V_n\}$ and $\{g_n\}$ generated by \eqref{eq:relatedVI1}-\eqref{eq:relatedVI3} converge to $V^*$ and $g^*$, respectively, as $n \to \infty$; the converged $V^*$ and $g^*$ satisfy the Bellman equation (see, e.g., Chapter 4 in \cite{bertsekas2012dynamic}).
	On this basis, we can prove the concavity of $V(\omega)$ by induction. Let us assume that $V_n(\omega)$ is concave. Then
	\begin{align*}
	{V_{n + 1}}(\omega ) = \min \left\{ {V_n^0(\omega ),V_n^1(\omega )} \right\},
	\end{align*}
	where
	\begin{align*}
	V_n^0(\omega ) &= H(\omega ) - g_{n+1} + {V_n}\left( {\tau (\omega )} \right),  \\
	V_n^1(\omega ) &= H(\omega ) - g_{n+1} + \omega {V_n}(1 - q) + (1 - \omega ){V_n}(p) + \lambda. 
	\end{align*}
	Since $H(\omega)$ and $V_n(\omega)$ are both concave, it is easy to verify that $V_n^0(\omega)$ and $V_n^1(\omega)$ are concave. Therefore, $V_{n+1}(\omega)$ is the minimum of two concave functions; hence it is also concave. Recall that we can set $V_0(\omega)=0$, then by induction, $V_n(\omega)$ is concave for all $n$. Since this recursion will eventually converge to the Bellman equation, we conclude that $V(\omega)$  is concave.	
\end{IEEEproof}

Based on the concavity of the value function, we next show that the optimal unichain policy for the single-bandit problem with a service charge $\lambda$ is a threshold policy. Without loss of generality, let $V(p)=0$. Define
\begin{align}
a(\omega,\lambda) & \triangleq \lambda + \omega V(1-q),  \\
r(\omega,\lambda) & \triangleq V(\tau(\omega)).
\end{align}
In the context that $\lambda$ is fixed, we will omit $\lambda$ and simply express the above two functions as $a(\omega)$ and $r(\omega)$. For a single-bandit problem with fixed $\lambda$, the Bellman equation can be written as
\begin{align}
V(\omega ) + g &= H(\omega ) + \min \{ a(\omega), r(\omega)\}. 
\end{align}
For any belief state $\omega \in [0,1]$, the active action is optimal for $\omega$ if $a(\omega)<r(\omega)$, otherwise the passive action is optimal. Note that $a(\omega)$ is a linear function of $\omega$. The threshold structure is obtained by proving the concavity of $r(\omega)$. 

\begin{proposition} \label{Props:1}
	The optimal policy for the single-bandit problem with a fixed service charge $\lambda$ is a threshold policy, for which there are two thresholds $0\le \omega^l\le \omega^u\le 1$ such that $a(\omega^l)=r(\omega^l)$, $a(\omega^u)=r(\omega^u)$, and that
	\begin{align*}
	u(t)=\begin{cases}
	1, &\text{if } \omega(t) \in (\omega^l,\omega^u)  \\
	0, &\text{otherwise.}
	\end{cases}
	\end{align*}
	The interval $(\omega^l,\omega^u)$ is referred to as the sampling region. If $(\omega^l,\omega^u)$ is empty, then the optimal policy is the never-sample policy, which takes passive action in all belief states.
\end{proposition}
\begin{IEEEproof}
	 Note that $a(\omega)$ is a linear function of $\omega$. Thus, to prove that the optimal policy is a threshold policy, it suffices to prove that $r(\omega)$  is concave. Since $r(\omega)=V(\tau(\omega))$, we have
	 \begin{align} \label{eq:prop1-1}
	 \frac{{{d^2}r(\omega )}}{{d{\omega ^2}}} = \frac{{{d^2}V(\tau )}}{{d{\tau ^2}}}{(1 - p - q)^2} < 0.
	 \end{align}
	 Inequality \eqref{eq:prop1-1} follows from the concavity of $V(\tau)$. Note that $r(\omega)$ is not guaranteed to be twice differentiable in all $\omega\in [0,1]$. Hence \eqref{eq:prop1-1} is only valid for $\omega$ for which $d^2 r/d\omega^2$ is well-defined. 
	 
	 Since $V(\omega)$ is concave, it is continuous, we then conclude that $r(\omega)=V(\tau)$ is continuous because $\tau = p+\omega (1-p-q)$ is also a continuous function of $\omega$.	 
	 For a particular $\hat{\omega}\in (0,1)$, let $\hat{\omega}^-$ and $\hat{\omega}^+$ denote the left-hand and right-hand limits of $\hat{\omega}$, respectively. Then the left-hand and right-hand derivatives of $r(\cdot)$ at $\hat{\omega}$ are
	 \begin{align}
	 &\frac{{dr({{\hat \omega }^ - })}}{{d\omega }}\triangleq \lim\limits_{\omega \to \hat \omega^-} \frac{{dr({ \omega  })}}{{d\omega }}   = \frac{{dV(\tau ({{\hat \omega }^- }))}}{{d\tau }}(1 - p - q),  \\
	 &\frac{{dr({{\hat \omega }^ + })}}{{d\omega }} \triangleq \lim\limits_{\omega \to \hat \omega^+} \frac{{dr({ \omega  })}}{{d\omega }}  = \frac{{dV(\tau ({{\hat \omega }^+ }))}}{{d\tau }}(1 - p - q).
	 \end{align}

	 To prove that $r(\omega)$ is concave even if $dr/d\omega$ is not continuous, we need to prove
	 \begin{align} \label{eq:prop1-2}
	 &\frac{{dr({{\hat \omega }^+ })}}{{d\omega }} - \frac{{dr({{\hat \omega }^- })}}{{d\omega }} =  \nonumber \\
	 &\left( {\frac{{dV(\tau ({{\hat \omega }^+ }))}}{{d\tau }} - \frac{{dV(\tau ({{\hat \omega }^- }))}}{{d\tau }}} \right)(1 - p - q) \le 0.
	 \end{align}
	 Let
	 \begin{align*}
	 \Delta d \triangleq  {\frac{{dV(\tau ({{\hat \omega }^+ }))}}{{d\tau }} - \frac{{dV(\tau ({{\hat \omega }^- }))}}{{d\tau }}} .
	 \end{align*}
	 If $p+q<1$, then $\tau(\hat \omega^+)>\tau(\hat \omega^-)$. We have $\Delta d\le0$ by the concavity of $V(\omega)$. Hence \eqref{eq:prop1-2} holds.
	 If $p+q>1$, then $\tau(\hat \omega^+)<\tau(\hat \omega^-)$. We have $\Delta d\ge0$ by the concavity of $V(\omega)$. Hence \eqref{eq:prop1-2} also holds. We thus conclude that $r(\omega)$ is concave. A concave function $r(\omega)$ and a linear function $a(\omega)$ have at most two intersections, say $\omega^l$ and $\omega^u$. Then $a(\omega)<r(\omega)$ if $\omega\in (\omega^l,\omega^u)$ and $a(\omega)\ge r(\omega)$ otherwise.  According to the Bellman equation, the active action is optimal for belief state $\omega$ if and only if  $a(\omega)<r(\omega)$. This completes the proof.	 
	 
\end{IEEEproof}

	Proposition 1 implies an interesting result. Since UoI may be a non-increasing function of the age of information, it is possible that the UoI of a single bandit decreases even if the associated remote process is not observed (i.e., under the passive action). A natural question in this case is that---as far as the single-bandit problem is concerned---is the active action still better than the passive action? We have the following result:
	\begin{corollary} \label{Coro:1}
		For any single-bandit problem with zero service charge (i.e., $\lambda=0$), the sampling region of the optimal policy is (0,1). 
	\end{corollary}
\begin{IEEEproof}
	This is an immediate result from Proposition \hyperref[Props:1]{1} and the observation that $a(0)=r(0)$ and $a(1)=r(1)$.
\end{IEEEproof}

Corollary 1 means that, if the service charge $\lambda = 0$, the active action is always better than the passive action. In other words, although UoI is not a monotonically increasing function of age, a specific remote process may still want to update its observation in every slot to minimize its average UoI. As $\lambda$ increases, we may expect that the passive action is better than the active action in more and more states, this is the desired property of indexability.
The threshold structure of the optimal policy for the single-bandit problem plays a crucial role in establishing the indexability. Since the sampling region $(\omega^l,\omega^u)$ is the set of belief states where the active action is optimal, the single-bandit is indexable if $(\omega^l,\omega^u)$ shrinks monotonically as $\lambda$ increases. We will prove the indexability in the next section. The following lemma contains properties that will be used in subsequent proofs.
\begin{lemma} \label{Lem:5}
	Let $(\omega^l,\omega^u)$ denote the sampling region of the optimal policy for the single-bandit problem with a service charge $\lambda$. Then for any $\lambda\ge 0$, we have the following:
	\begin{itemize}
		\item[1.] If $\omega^* \notin (\omega^l,\omega^u)$, then $g=H(\omega^*)$.
		\item [2.] $\omega^u\ge \omega^*$.
	\end{itemize}
\end{lemma}
\begin{IEEEproof}
	Statement 1 follows directly from the Bellman equation:
	\begin{align*}
	V({\omega ^ * }) + g = H({\omega ^ * }) + V\left( {\tau ({\omega ^ * })} \right) = H({\omega ^ * }) + V\left( {{\omega ^ * }} \right).
	\end{align*}
	Statement 2 means that, if $\omega^* \notin(\omega^l,\omega^u)$, then the only possibility is $\omega^* \le \omega^l\le \omega^u$. This result is established based on our assumption that $p\le q$. See Appendix \hyperref[Apd:A]{A} for the complete proof.
\end{IEEEproof}

\section{Indexability of the RMAB}  \label{sec:index}
This section applies the properties established in Section IV to prove that the single-bandit problem is indexable. As stated in Lemma \hyperref[Lem:1]{1}, monotonic bandits ($p+q<1$) and oscillating bandits ($p+q>1$) have different characteristics. Hence we will establish indexability for the two cases separately.

According to Proposition \hyperref[Props:1]{1}, the optimal policy for the single-bandit problem with any service charge $\lambda$  has a continuous sampling region $(\omega^l,\omega^u)$. Note that if $(\omega^l,\omega^u)=(0,1)$, it is equally optimal to take active actions in the entire belief state space $[0,1]$. Therefore, to prove the bandit is indexable, it suffices to show that $(\omega^l,\omega^u)$ shrinks monotonically from $(0,1)$ to null as $\lambda$ increases. The lemma below presents a sufficient condition for this monotonicity.
\begin{lemma}[Monotonicity Condition]  \label{Lem:6}
	Let $(\omega^l,\omega^u)$ denote the sampling region of the optimal policy for the single-bandit problem with a service charge $\lambda$. Suppose that for any $\lambda\ge 0$ such that $(\omega^l,\omega^u)$ is non-empty, we have
	\begin{align} \label{eq:monoteq}
	\frac{{\partial a({\omega ^l},\lambda )}}{{\partial \lambda }} > \frac{{\partial r({\omega ^l},\lambda )}}{{\partial \lambda }}\text{ and }\frac{{\partial a({\omega ^u},\lambda )}}{{\partial \lambda }} > \frac{{\partial r({\omega ^u},\lambda )}}{{\partial \lambda }}.
	\end{align}
	Then $\omega^l$ monotonically increases with $\lambda$, while $\omega^u$ monotonically decreases with $\lambda$. Hence $(\omega^l,\omega^u)$ shrinks monotonically from $(0,1)$ toward the empty set as $\lambda$ increases from 0 to $\infty$. Condition \eqref{eq:monoteq} is referred to as the monotonicity condition.
\end{lemma}
\begin{IEEEproof}
	We prove this lemma by contradiction. Assume the monotonicity condition is met and there exits a $\lambda\ge 0$ such that $\omega^l$ is decreasing at $\lambda$. Then there exists a $\Delta >0$ such that for any $\delta \in [0,\Delta]$, we have
	\begin{align}
	a({\omega ^l},\lambda  + \delta ) \le r({\omega ^l},\lambda  + \delta ).
	\end{align}
	Note that $a({\omega ^l},\lambda ) = r({\omega ^l},\lambda )$, we thus have
	\begin{align*}
	\frac{{\partial a({\omega ^l},\lambda )}}{{\partial \lambda }} = \mathop {\lim }\limits_{\delta  \to 0} \frac{{a({\omega ^l},\lambda  + \delta ) - a({\omega ^l},\lambda )}}{\delta } \\
	\le \mathop {\lim }\limits_{\delta  \to 0} \frac{{r({\omega ^l},\lambda  + \delta ) - r({\omega ^l},\lambda )}}{\delta } = \frac{{\partial r({\omega ^l},\lambda )}}{{\partial \lambda }}{\rm{  }},
	\end{align*}
	which contradicts \eqref{eq:monoteq}. We thus conclude that $\omega^l$ monotonically increases with $\lambda$ if the monotonicity condition is met. Applying a similar argument can prove that $\omega^u$ monotonically decreases with $\lambda$ if the monotonicity condition is met. Recall that when $\lambda=0$, the thresholds of the optimal policy are $\omega^l=0$ and $\omega^u=1$. Hence $(\omega^l,\omega^u)$ shrinks from the entire state space $(0,1)$ to the empty set as $\lambda$ increases from 0 to $\infty$ if \eqref{eq:monoteq} holds for all $\lambda\ge 0$. 
\end{IEEEproof}

Applying Lemma 6, we can establish the indexability by examining the partial derivatives of $a(\omega,\lambda)$ and $r(\omega,\lambda)$ w.r.t. $\lambda$. Clear expressions of $a(\omega,\lambda)$ and $r(\omega,\lambda)$ require the expression of $V(\omega)$. In the remaining part of this section, we first derive the expression of $V(\omega)$, and then apply the monotonicity condition to establish the indexability of our UoI scheduling RMAB.

\subsection{Hitting Time}
To obtain the expression of $V(\omega)$, we are interested in the hitting time for the bandit to enter a belief state in the sampling region starting with any initial belief state. The formal definition of hitting time is given below:
\begin{definition}[Hitting time]
	For the optimal threshold policy with sampling region $(\omega^l,\omega^u)$, the hitting time of belief state $\omega\in [0,1]$ is defined as the minimum number of slots required for the bandit to evolve from $\omega$ to a belief state in $(\omega^l,\omega^u)$. In particular, let $T(\omega,\omega^l,\omega^u)$ denote the hitting time of $\omega$. Then
	\begin{align*}
	T(\omega ,{\omega ^l},{\omega ^u}) \triangleq \min \left\{ {k:{\tau ^k}(\omega ) \in \left( {{\omega ^l},{\omega ^u}} \right),k = 0,1,2,...} \right\},
	\end{align*}
	where $\tau^0(\omega)\triangleq \omega$.
\end{definition}

For any belief state $\omega$, if the hitting time of $\omega$ is $T(\omega ,{\omega ^l},{\omega ^u})=m<\infty$, then the optimal policy takes the passive action in belief states $\{ {\tau ^k}(\omega ),k = 0, \cdots ,m - 1\} $ and the active action in belief state $\tau^m(\omega)$. Hence the value function can be written as follows:
\begin{align}
V(\omega ) = \sum\limits_{k = 0}^m {\left[ {H({\tau ^k}(\omega )) - g} \right]}  + {\tau ^m}(\omega )V(1 - q) + \lambda. 
\end{align}
Monotonic bandits ($p+q<1$) and oscillating bandits ($p+q>1$ ) have different expressions of $T(\omega ,{\omega ^l},{\omega ^u})$. We need to discuss the two cases separately. The lemma below presents the hitting time for monotonic bandits.
\begin{lemma}  \label{Lem:7}
	Let $(\omega^l,\omega^u)$ denote the sampling region of the optimal policy for a monotonic bandit ($p+q<1$) with a service charge $\lambda$. The hitting time $T(\omega ,{\omega ^l},{\omega ^u})$ of any $\omega\in [0,1]$ is given by
	\begin{align*}
	T(\omega ,{\omega ^l},{\omega ^u}) = \begin{cases}
	0,\qquad \qquad \qquad \qquad \quad \ \text{ if }{\omega ^l} < \omega  < {\omega ^u}\\
	\left\lfloor {{{\log }_{1 - p - q}}\frac{{{\omega ^l} - {\omega ^ * }}}{{\omega  - {\omega ^ * }}}} \right\rfloor  + 1,\quad \text{if }\omega  \le {\omega ^l} < {\omega ^ * }\\
	{k_0} \triangleq \left\lfloor {{{\log }_{1 - p - q}}\frac{{{\omega ^u} - {\omega ^ * }}}{{\omega  - {\omega ^ * }}}} \right\rfloor + 1, \\
	\qquad \qquad \text{ if }\omega  \ge {\omega ^u} > {\omega ^ * },{\tau ^{{k_0}}}(\omega ) > {\omega ^l} \\
	\infty,\qquad \qquad \qquad \qquad \quad \ \text{otherwise }
	\end{cases}
	\end{align*}
	where $\omega^* = p/(p+q)$ is the equilibrium belief state. 
\end{lemma}
\begin{IEEEproof}
	First, if $\omega\in (\omega^l,\omega^u)$, then $T(\omega ,{\omega ^l},{\omega ^u})=0$ by definition. If $\omega\notin (\omega^l,\omega^u)$, the bandit governed by a threshold policy will keep taking the passive action for  $T(\omega ,{\omega ^l},{\omega ^u})$ slots until it evolves to a belief state in $ (\omega^l,\omega^u)$. According to Lemma \hyperref[Lem:5]{5}, we have $\omega^u \ge \omega^*$. Hence there are three cases for $\omega\notin (\omega^l,\omega^u)$: (1) $\omega  \ge {\omega ^u} \ge {\omega ^ * }$; (2) $\omega  \le {\omega ^l} < {\omega ^ * }$; (3) $\omega  \le {\omega ^l} \ge {\omega ^ * }$.
	
	As stated in Lemma \hyperref[Lem:1]{1}, after consecutively taking the passive action for $k$ slots, the belief state of a monotonic bandit evolves from $\omega$ to $\tau^k(\omega)$, and
	\begin{align} \label{eq:Lemma7-1}
	\frac{{{\tau ^k}(\omega ) - {\omega ^ * }}}{{\omega  - {\omega ^ * }}} = {(1 - p - q)^k}.
	\end{align}
	If $\omega  \ge {\omega ^u} \ge {\omega ^ * }$, then $\tau^k(\omega)$ monotonically decreases as $k$ increases. The hitting time is equal to the integer $k$ such that ${\tau ^{k - 1}}(\omega ) \ge {\omega ^u} > {\tau ^k}(\omega )$. For \eqref{eq:Lemma7-1}, there exists a real value $x$ such that ${\omega ^u} - {\omega ^ * } = (\omega  - {\omega ^ * }){(1 - p - q)^x}$. Then the hitting time is 
	\begin{align*}
	T(\omega ,{\omega ^l},{\omega ^u}) = k = \left\lfloor x \right\rfloor  + 1 = \left\lfloor {{{\log }_{1 - p - q}}\frac{{{\omega ^u} - {\omega ^ * }}}{{\omega  - {\omega ^ * }}}} \right\rfloor  + 1.
	\end{align*}
	Note that there may exist some $\lambda$ such that ${\omega ^u} > {\omega ^l} > {\omega ^ * }$. If this is the case and ${\tau ^k}(\omega ) \le {\omega ^l}$, then $T(\omega ,{\omega ^l},{\omega ^u})=\infty$ because ${\tau ^k}(\omega ) \notin ({\omega ^l},{\omega ^u})$ for all $k$. 
	
	If $\omega  \le {\omega ^l} < {\omega ^ * }$, then $\tau^k(\omega)$ monotonically increases as $k$ increases. The hitting time is equal to the integer $k$ such that ${\tau ^{k - 1}}(\omega ) \le {\omega ^l} < {\tau ^k}(\omega )$. Similarly, we can obtain the hitting time:
	\begin{align*}
	T(\omega ,{\omega ^l},{\omega ^u}) = \left\lfloor {{{\log }_{1 - p - q}}\frac{{{\omega ^l} - {\omega ^ * }}}{{\omega  - {\omega ^ * }}}} \right\rfloor  + 1.
	\end{align*}
	
	Finally, if $\omega  \le {\omega ^l} \ge {\omega ^ * }$, then $\tau^k(\omega)<\omega^l$ for all $k$. This means that $\tau^k(\omega)\notin (\omega^l,\omega^u)$ for all $k$. Hence $T(\omega ,{\omega ^l},{\omega ^u})=\infty$.
\end{IEEEproof}
	
The hitting time of an oscillating bandit is presented in the following lemma:
\begin{lemma}  \label{Lem:8}
	Let $(\omega^l,\omega^u)$ denote the sampling region of the optimal policy for an oscillating bandit ($p+q>1$) with a service charge $\lambda$. Then,
	\begin{itemize}
		\item [1. ] If $\omega^*\in [\omega^l,\omega^u]$, the hitting time $T(\omega ,{\omega ^l},{\omega ^u})$ is given by \eqref{eq:Lemma8-1}. 
		\item [2. ] If $\omega^*\notin [\omega^l,\omega^u]$, the hitting time $T(\omega ,{\omega ^l},{\omega ^u})$ is given by \eqref{eq:Lemma8-2}.
	\end{itemize}
	
	\newcounter{TempEqCnt} 
	\setcounter{TempEqCnt}{\value{equation}} 
	\setcounter{equation}{32} 
	\begin{figure*}[ht] 
		\begin{align} \label{eq:Lemma8-1}
		&T(\omega ,{\omega ^l},{\omega ^u}) = \begin{cases}
		\min \left\{ {2\left\lfloor {\phi ({\omega ^u},\omega )} \right\rfloor  + 3,{\kern 1pt} {\kern 1pt} {\kern 1pt} {\kern 1pt} 2\left\lfloor {\varphi ({\omega ^l},\omega )} \right\rfloor  + 2} \right\},&\text{if }\omega  \le {\omega ^l}\\
		0,&\text{if }{\omega ^l} < \omega  < {\omega ^u}\\
		\min \left\{ {2\left\lfloor {\phi ({\omega ^l},\omega )} \right\rfloor  + 3,{\kern 1pt} {\kern 1pt} {\kern 1pt} {\kern 1pt} {\kern 1pt} 2\left\lfloor {\varphi ({\omega ^u},\omega )} \right\rfloor  + 2} \right\},&\text{if }\omega  \ge {\omega ^u}
		\end{cases}   \\  \label{eq:Lemma8-2}
		&T(\omega ,{\omega ^l},{\omega ^u}) = \begin{cases}
		0,&\text{if }\omega  \in ({\omega ^l},{\omega ^u})\\
		{k_1} \triangleq 2\left\lfloor {\varphi ({\omega ^u},\omega )} \right\rfloor  + 2,&\text{if }\omega  \ge {\omega ^u}\text{ and }{k_1} \le 2\left\lceil {\varphi ({\omega ^l},\omega )} \right\rceil  - 2\\
		{k_2} \triangleq 2\left\lfloor {\phi ({\omega ^u},\omega )} \right\rfloor  + 3,&\text{if }\omega  \le {\omega ^l},\tau (\omega ) > {\omega ^l}\text{ and }{k_2} \le 2\left\lceil {\phi ({\omega ^l},\omega )} \right\rceil  - 1\\
		\infty ,&\text{otherwise}
		\end{cases} \\ \notag
		\text{where   }\quad &	\\  \notag
		& \varphi (x,y) \triangleq  \frac{1}{2}{\log _{p + q - 1}}\frac{{x - {\omega ^ * }}}{{y - {\omega ^ * }}}, \quad
		\phi (x,y) \triangleq \frac{1}{2}{\log _{p + q - 1}}\frac{{x - {\omega ^ * }}}{{{\omega ^ * } - y}} - \frac{1}{2}.	 
		\end{align}
		\hrulefill  
	\end{figure*}

\end{lemma}
\begin{IEEEproof}
	The basic idea of this proof is similar to the proof of Lemma \hyperref[Lem:7]{7}. However, since $p+q>1$, as stated in Lemma 1, $\tau^k(\omega)$ is an oscillating function of $k$. In particular, there are two branches, $\tau^{2k}(\omega)$ and $\tau^{2k+1}(\omega)$, that approach to $\omega^*$ from opposite directions. Hence the expression of $T(\omega ,{\omega ^l},{\omega ^u})$ for oscillating bandits is more complicated than monotonic bandits. See Appendix \hyperref[apd:C]{B} for the complete proof.
\end{IEEEproof}

\subsection{Indexability}
With the monotonicity condition and the closed-form expression of the value function based on the hitting time, we can prove that monotonic bandits ($p+q<1$) and oscillating bandits ($p+q>1$) are indexable.

\begin{theorem}
	The RMAB problem P1 is indexable.
\end{theorem}
\begin{IEEEproof}
	This proof is divided into two parts. We prove in Proposition \hyperref[Props:2]{2} that any monotonic bandit is indexable, and then prove in Proposition \hyperref[Props:3]{3} that any oscillating bandit is indexable. Since all bandits are indexable, the RMAB is indexable. See the two propositions below for details.  
\end{IEEEproof}

\begin{figure*}[!t]
	\centering
	\begin{subfigure}[b]{0.32\linewidth}
		\includegraphics[width=\linewidth]{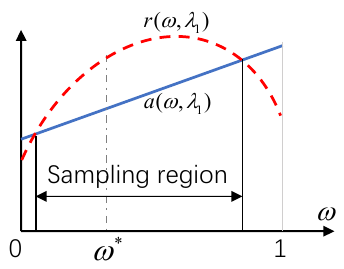}
		\caption{service charge $\lambda_1$: $\omega^* \in (\omega^l, \omega^u)$.}
		\label{fig:sampleR1}
	\end{subfigure}
	\begin{subfigure}[b]{0.32\linewidth}
		\includegraphics[width=\linewidth]{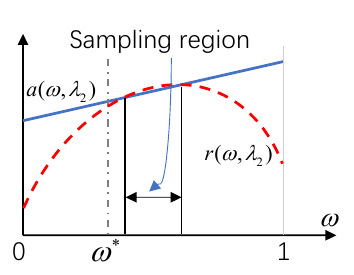}
		\caption{service charge $\lambda_2$: $\omega^* \notin (\omega^l, \omega^u)$.}
		\label{fig:sampleR2.pdf}
	\end{subfigure}
	\begin{subfigure}[b]{0.32\linewidth}
		\includegraphics[width=\linewidth]{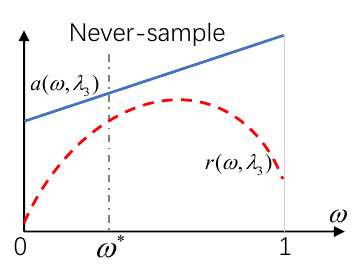}
		\caption{service charge $\lambda_3$: $(\omega^l, \omega^u)$ is null.}
		\label{fig:sampleR3.pdf}
	\end{subfigure}
	\caption{Sampling region of the optimal policy for the single-bandit problem under different service charges, $\lambda_1< \lambda_2 <\lambda_3$.}
	\label{fig:sampleregion}
\end{figure*}

We first establish the indexability for monotonic bandits. Basically, we examine the partial derivatives of $a(\omega,\lambda)$ and $r(\omega,\lambda)$ w.r.t. $\lambda$ to show that the monotonicity condition holds for all $\lambda\ge 0$. Fig. \ref{fig:sampleregion} illustrates how the optimal sampling region varies with $\lambda$.
At first, $(\omega^l, \omega^u)$ includes the equilibrium belief state $\omega^*$. If the monotonicity condition holds in this case, then  $(\omega^l, \omega^u)$ shrinks as $\lambda$ increases, and we may arrive at another case that $\omega^* \notin (\omega^l, \omega^u)$. According to Lemma \hyperref[Lem:7]{7}, $T(\omega ,{\omega ^l},{\omega ^u})$ has different forms in these two cases.
In Proposition 2, we prove that the monotonicity condition holds for both the two cases. Consequently, $(\omega^l, \omega^u)$ would shrink to the empty set when $\lambda$ is large enough.

\begin{proposition} \label{Props:2}
	A monotonic bandit ($p+q<1$) is indexable.
\end{proposition}
\begin{IEEEproof}
	To establish the indexability of a monotonic bandit, according to Lemma \hyperref[Lem:6]{6}, it suffices to show that the monotonicity condition holds for all $\lambda\ge0$. Recall that $(\omega^l, \omega^u)=(0,1)$ when $\lambda=0$. We first show that the monotonicity condition holds if $\omega^* \in (\omega^l, \omega^u)$, and then show it still holds if $\omega^* \notin (\omega^l, \omega^u)$.
	
	First, if $\omega^* \in (\omega^l, \omega^u)$, then $T(p ,{\omega ^l},{\omega ^u})<\infty$ and $T(1-q ,{\omega ^l},{\omega ^u})<\infty$ by Lemma \hyperref[Lem:7]{7}. Note that ${\tau ^k}(p) = {p^{(k + 1)}}$. For notation simplicity, let $L \triangleq T(p,{\omega ^l},{\omega ^u}) + 1$ and $K \triangleq T(1-q,{\omega ^l},{\omega ^u}) + 1$. Then according to the Bellman equation, we have
	\begin{align} \label{eq:Prop2-1}
	&V(p) = \sum\limits_{k = 1}^L {\left[ {H\left( {{p^{(k)}}} \right) - g} \right]}  + \lambda  + {p^{(L)}}V(1 - q), \\  \label{eq:Prop2-2}
	&V(1 - q) = \sum\limits_{k = 1}^K {\left[ {H\left( {{q^{(k)}}} \right) - g} \right]}  + \lambda  + \left( {1 - {q^{(K)}}} \right)V(1 - q).
	\end{align}
	Since we set $V(p)=0$, it follows from \eqref{eq:Prop2-1} and \eqref{eq:Prop2-2} that
	\begin{align}  \label{eq:Prop2-vq}
	&V(1 - q) = \frac{1}{{{q^{(K)}}}}\left( {\sum\limits_{k = 1}^K {\left[ {H\left( {{q^{(k)}}} \right) - g} \right]}  + \lambda } \right), \\ \label{eq:Prop2-g}
	&g = \frac{{{q^{(K)}}\left[ {\sum\limits_{k = 1}^L {H\left( {{p^{(k)}}} \right)}  + \lambda } \right] + {p^{(L)}}\left[ {\sum\limits_{k = 1}^K {H\left( {{q^{(k)}}} \right)}  + \lambda } \right]} }{{L{q^{(K)}} + K{p^{(L)}}}}.
	\end{align}
	For $a(\omega ,\lambda ) = \omega V(1 - q) + \lambda $, its partial derivative w.r.t. $\lambda$ is given by
	\begin{align*}
	\frac{{\partial a(\omega ,\lambda )}}{{\partial \lambda }} = \frac{\omega }{{{q^{(K)}}}}\left( {1 - K\frac{{\partial g}}{{\partial \lambda }}} \right) + 1 = \frac{{\omega \left( {L - K} \right)}}{{L{q^{(K)}} + K{p^{(L)}}}} + 1.
	\end{align*}	
	Since $\omega^* \in (\omega^l, \omega^u)$ and $p+q<1$, we have $ T(\omega^l,{\omega ^l},{\omega ^u})=1$ and $ T(\omega^u,{\omega ^l},{\omega ^u})=1$. Then for $\omega= \omega^l$ or $\omega^u$, the expression of $r(\omega,\lambda)$ is
	\begin{align}
	r(\omega ,\lambda ) = V\left( {\tau (\omega )} \right) = H\left( {\tau (\omega )} \right) - g + a\left( {\tau (\omega ),\lambda } \right).
	\end{align}
	It follows that
	\begin{align}
	\frac{{\partial r(\omega ,\lambda )}}{{\partial \lambda }} = \frac{{\tau (\omega )\left( {L - K} \right) - \left( {{q^{(K)}} + {p^{(L)}}} \right)}}{{L{q^{(K)}} + K{p^{(L)}}}} + 1.
	\end{align}
	To verify the monotonicity condition, define
	\begin{align*}
	{f_\delta }(\omega ) &\triangleq \left( {L{q^{(K)}} + K{p^{(L)}}} \right)\left( {\frac{{\partial r(\omega ,\lambda )}}{{\partial \lambda }} - \frac{{\partial a(\omega ,\lambda )}}{{\partial \lambda }}} \right) \\
	&= (L - K)\left[ {p - \omega (p + q)} \right] - {q^{(K)}} - {p^{(L)}}.
	\end{align*}
	It can be proved that ${f_\delta }(\omega^l )<0$ and ${f_\delta }(\omega^u )<0$ (see Lemma C1 in Appendix \hyperref[apd:D]{C}). Therefore, the monotonicity condition holds when $\omega^* \in (\omega^l, \omega^u)$; this means that $(\omega^l, \omega^u)$ shrinks from $(0,1)$ as $\lambda$ increase from 0.
	
	Assume there is a $\lambda^*$ such that we arrive at the case of $\omega^l=\omega^*$ and/or $\omega^u=\omega^*$. If $\omega^l$ and $\omega^u$ are both equal to $\omega^*$, then $(\omega^l, \omega^u)$ is null; hence the indexability is already established. Next, we focus on the case that $\omega^l\neq \omega^u$ when $\lambda = \lambda^*$. Recall that $\omega^u$ can not be smaller than $\omega^*$. We therefore have two cases for $\lambda \ge \lambda^*$: (i) ${\omega ^u} > {\omega ^l}\; \ge {\omega ^ * }$; (ii) ${\omega ^u} = {\omega ^ * } > {\omega ^l}$. We first establish the monotonicity condition in case (i), and then show that case (ii) is impossible.
	
	(i) If there exists a $\lambda$ such that ${\omega ^u} > {\omega^l} \ge {\omega^*}$, then according to the Bellman equation, the expression of $V(1-q)$ is the same as given by \eqref{eq:Prop2-vq}; while $V(p)$ is given as follows:
	\begin{align}
	V(p) = \mathop {\lim }\limits_{L \to \infty } \left\{ {\sum\limits_{k = 1}^L {\left[ {H\left( {{p^{(k)}}} \right) - g} \right]}  + V\left( {{p^{(L)}}} \right)} \right\} = 0.
	\end{align}
	Since $\mathop {\lim }\limits_{k \to \infty } {p^{(k)}} = {\omega ^ * } = \mathop {\lim }\limits_{k \to \infty } {\tau ^k}({\omega ^l})$, we obtain
	\begin{align} \notag
	&r({\omega ^l},\lambda ) = \mathop {\lim }\limits_{k \to \infty } \left\{ {\sum\limits_{i = 1}^k {\left[ {H\left( {{\tau ^i}({\omega ^l})} \right) - g} \right]}  + V\left( {{\tau ^k}({\omega ^l})} \right)} \right\}\\ \label{eq:Prop2-rw}
	=& \mathop {\lim }\limits_{k \to \infty } \sum\limits_{i = 1}^k {\left[ {H\left( {{\tau ^i}({\omega ^l})} \right) - g} \right]}  - \mathop {\lim }\limits_{L \to \infty } \sum\limits_{i = 1}^L {\left[ {H\left( {{p^{(i)}}} \right) - g} \right]}. 
	\end{align}
	As stated in Lemma \hyperref[Lem:5]{5}, the optimal average penalty $g=H(\omega^*)$ if $\omega^* \notin (\omega^l, \omega^u)$; hence $\partial g/\partial \lambda =0$. Then 
	\begin{align} \label{eq:Prop2-43}
	\frac{{\partial a({\omega ^l},\lambda )}}{{\partial \lambda }}  = \frac{{{\omega ^l}}}{{{q^{(K)}}}} + 1 > \frac{{\partial r({\omega ^l},\lambda )}}{{\partial \lambda }} = 0.
	\end{align}
	
	For $\omega^u$, we have two cases: 1) $\tau(\omega^u)>\omega^l$; 2) $\tau(\omega^u)\le \omega^l$. First, if $\tau(\omega^u)>\omega^l$, then the optimal policy takes the active action in belief state $\tau(\omega^u)$. We have
	\begin{align*}
	r({\omega ^u},\lambda ) = V(\tau ({\omega ^u})) = H(\tau ({\omega ^u})) - g + \tau ({\omega ^u})V(1 - q) + \lambda. 
	\end{align*}
	Hence
	\begin{align} \label{eq:Prop2-44}
	\frac{{\partial r({\omega ^u},\lambda )}}{{\partial \lambda }} = \frac{{\tau ({\omega ^u})}}{{{q^{(K)}}}} + 1 < \frac{{{\omega ^u}}}{{{q^{(K)}}}} + 1 = \frac{{\partial a({\omega ^u},\lambda )}}{{\partial \lambda }}.
	\end{align}	
	The inequality follows from $\tau(\omega^u)<\omega^u$.
	
	On the other hand, if $\tau(\omega^u)\le\omega^l$, then the optimal policy takes the passive action in belief states $\{\tau^k(\omega^u)\}$ for all $k$. In this case, $r(\omega^u,\lambda)$ is of the same form as  $r(\omega^l,\lambda)$ given by \eqref{eq:Prop2-rw}. Therefore,
	\begin{align} \label{eq:Prop2-45}
	\frac{{\partial r({\omega ^u},\lambda )}}{{\partial \lambda }} = 0 <\frac{{{\omega ^u}}}{{{q^{(K)}}}} + 1 = \frac{{\partial a({\omega ^u},\lambda )}}{{\partial \lambda }}.  
	\end{align}
	Inequalities \eqref{eq:Prop2-43}-\eqref{eq:Prop2-45} show that the monotonicity condition holds in the case of ${\omega ^u} > {\omega^l} \ge {\omega^*}$. Note that we assume $K<\infty$ in this case. According to Lemma \hyperref[Lem:7]{7}, $K = T(1-q,{\omega ^l},{\omega ^u}) + 1$ may be infinite if $\omega^u$ and $\omega^l$ are close enough such that $1 - {q^{(k)}} \notin ({\omega ^l},{\omega ^u})$ for all $k$. If this is the case, we argue that the indexability is already established because $(\omega^l, \omega^u)$ does not contain any belief state in $\Omega  = \{ {p^{(n)}},1 - {q^{(n)}}:n \in {\mathbb{N}^ + }\} $. Recall that the real belief state space of the bandit is $\Omega$; we extend the state space to $[0,1]$ just to ease analysis. In fact, it turns out that the monotonicity condition still holds in this case, but we omit the derivation for simplicity. 
	
	(ii) If ${\omega ^u} = {\omega ^ * } > {\omega ^l}$. The expression of $V(p)$ is the same as given by \eqref{eq:Prop2-1}. Since $V(p)=0$ and $K=\infty$, it follows from \eqref{eq:Prop2-1} that
	\begin{align*}
	V(1 - q) &=  - \frac{1}{{{p^{(L)}}}}\left( {\sum\limits_{i = 1}^L {\left[ {H\left( {{p^{(i)}}} \right) - g} \right]}  + \lambda } \right) \\
	&= \mathop {\lim }\limits_{K \to \infty } \left\{ {\sum\limits_{i = 1}^K {\left[ {H\left( {{q^{(i)}}} \right) - g} \right]}  + V\left( {1 - {q^{(K)}}} \right)} \right\}.
	\end{align*}
	Note that $\mathop {\lim }\limits_{k \to \infty } \left( {1 - {q^{(k)}}} \right) = {\omega ^ * } = {\omega ^u}$ and $g=H(\omega^*)$, then
	\begin{align*}
	r({\omega ^u},\lambda ) = V({\omega ^ * }) = V(1 - q) - \mathop {\lim }\limits_{k \to \infty } \sum\limits_{i = 1}^k {\left[ {H\left( {{q^{(i)}}} \right) - g} \right]}, 
	\end{align*}
	and
	\begin{align} \label{eq:Prop2-46}
	\frac{{\partial r({\omega ^u},\lambda )}}{{\partial \lambda }}  =  - \frac{1}{{{p^{(L)}}}} < \frac{{\partial a({\omega ^u},\lambda )}}{{\partial \lambda }} =  - \frac{{{\omega ^ * }}}{{{p^{(L)}}}} + 1.
	\end{align}
	According to Lemma \hyperref[Lem:6]{6}, inequality \eqref{eq:Prop2-46} means that $\omega^u$ would decrease as $\lambda$ increases. Hence $\omega^u<\omega^*$ when $\lambda>\lambda^*$, which contradicts Lemma \hyperref[Lem:5]{5}. Therefore,  ${\omega ^u} = {\omega ^ * } > {\omega ^l}$ is an impossible case.
	
	In summary, we conclude that the monotonicity condition holds whenever $(\omega^l, \omega^u)$ is nonempty. This means that $(\omega^l, \omega^u)$ shrinks from the entire state space to the empty set as $\lambda$ increases from 0 to $\infty$; we thus establish the indexability for monotonic bandits.
\end{IEEEproof}

For monotonic bandits, any threshold policy is unichain. However, oscillating bandits do not share the property. To see this, consider a threshold policy with sampling region $(\omega^l, \omega^u)$, where $\omega^l>\omega^*$.
For monotonic bandits, the threshold policy is unichain because ${p^{(n)}} \notin ({\omega ^l},{\omega ^u})$ for all $n$ (according to Lemma \hyperref[Lem:2]{2}). For oscillating bandits, on the other hand, we may have some odd integers $n$ and even integers $m$ such that ${p^{(n)}},1 - {q^{(m)}} \in ({\omega ^l},{\omega ^u})$; hence the threshold policy is multichain. We thus need a slightly more careful argument in the case of oscillating bandits. As will be shown in the proof of Proposition 3, for an oscillating bandit, there may exist a service charge $\lambda^*$ such that some multichain policies are optimal. Meanwhile, there is a unichain policy that is equally optimal, as stated in Lemma \hyperref[Lem:3]{3}. In this case, each of these optimal policies has a different sampling region, hence multiple belief states---thresholds of these equally optimal sampling regions---share the same Whittle index.

\begin{proposition} \label{Props:3}
	An oscillating bandit ($p+q>1$) is indexable. In addition, if there is a $\lambda^*$ such that the optimal sampling region is $({\omega ^l} , {\omega ^u})$, where  ${\omega ^l} = {\omega ^ * } < {\omega ^u}$, then the Whittle index of every belief state in $[{\omega ^ * },\tilde \omega ]$ is $\lambda^*$, where
	\begin{align*}
	\tilde \omega  \triangleq \min \big\{&\max \{ {p^{(k)}}:{p^{(k)}} < {\omega ^u},k \in {\mathbb{N}^ + }\} , \\
	&\max \{ 1 - {q^{(k)}}:1 - {q^{(k)}} < {\omega ^u},k \in {\mathbb{N}^ + }\} \big\}. 
	\end{align*}
\end{proposition}
\begin{IEEEproof}
	We can use a similar method as in Proposition \hyperref[Props:2]{2} to establish indexability for oscillating bandits. We first establish the monotonicity condition for the case of $\omega^*\in(\omega^l, \omega^u)$. Then we analyze the case of $\omega^*\notin(\omega^l, \omega^u)$. Also, let $L \triangleq T(p,{\omega ^l},{\omega ^u}) + 1$ and $K \triangleq T(1-q,{\omega ^l},{\omega ^u}) + 1$.
	
	First, if $\omega^*\in(\omega^l, \omega^u)$, then $L<\infty$ and $K<\infty$. The expressions of $V(p)$ and $V(1-q)$ are the same as given by \eqref{eq:Prop2-1} and \eqref{eq:Prop2-2} in the proof of Proposition 2. Hence we can obtain the partial derivative of $a(\omega ,\lambda ) = \omega V(1 - q) + \lambda $ w.r.t. $\lambda$ as follows:
	\begin{align} \label{eq:Prop3-da}
	\frac{{\partial a(\omega ,\lambda )}}{{\partial \lambda }}  = \frac{{\omega \left( {L - K} \right)}}{{L{q^{(K)}} + K{p^{(L)}}}} + 1.
	\end{align}
	Given that $\omega^l <\omega^*$, $\tau(\omega^l)$ may not belong to $({\omega ^l} , {\omega ^u})$ because it is possible that $\tau(\omega^l)>\omega^u$. However, we must have $\tau^2(\omega^l)\in({\omega ^l} , {\omega ^u})$ because $\omega^l<\tau^2(\omega^l)<{\omega ^*} $. Hence $T(\omega^l,{\omega ^l} , {\omega ^u})=1$ or 2. Similarly, $T(\omega^u,{\omega ^l} , {\omega ^u})=1$ or 2. Therefore, for $\omega=\omega^l$ or $\omega^u$, function $r(\omega,\lambda)$ has two possible forms:
	\begin{align*}
	r(\omega ,\lambda ) = \begin{cases}
	H\left( {\tau (\omega )} \right) - g + a\left( {\tau (\omega ),\lambda } \right), \ \text{if }T(\omega ,{\omega ^l},{\omega ^u}) = 1\\
	\sum\limits_{k = 1}^2 {\left[ {H\left( {{\tau ^k}(\omega )} \right) - g} \right]}  + a\left( {{\tau ^2}(\omega ),\lambda } \right), \\
	\qquad \qquad \qquad \qquad \qquad \qquad \text{if }T(\omega ,{\omega ^l},{\omega ^u}) = 2
	\end{cases}
	\end{align*}
	Then if $T(\omega ,{\omega ^l},{\omega ^u}) = 1$,
	\begin{align} \label{eq:Prop3-dr1}
	\frac{{\partial r(\omega ,\lambda )}}{{\partial \lambda }} = 
	\frac{{\tau (\omega )\left( {L - K} \right) - \left( {{q^{(K)}} + {p^{(L)}}} \right)}}{{L{q^{(K)}} + K{p^{(L)}}}} + 1.	
	\end{align}
	While if $T(\omega ,{\omega ^l},{\omega ^u}) = 2$,
	\begin{align}  \label{eq:Prop3-dr2}
	\frac{{\partial r(\omega ,\lambda )}}{{\partial \lambda }} = 
	\frac{{{\tau ^2}(\omega )\left( {L - K} \right) - 2\left( {{q^{(K)}} + {p^{(L)}}} \right)}}{{L{q^{(K)}} + K{p^{(L)}}}} + 1.	
	\end{align}
	To verify the monotonicity condition, we need to compare the partial derivatives of $a(\omega,\lambda)$ and $r(\omega,\lambda)$. For the case of $T(\omega ,{\omega ^l},{\omega ^u}) = 1$, define
	\begin{align*}
	{f_1}(\omega ) &\triangleq  \left( {L{q^{(K)}} + K{p^{(L)}}} \right)\left( {\frac{{\partial r(\omega ,\lambda )}}{{\partial \lambda }} - \frac{{\partial a(\omega ,\lambda )}}{{\partial \lambda }}} \right) \\
	&= (L - K)\left[ {p - \omega (p + q)} \right] - {q^{(K)}} - {p^{(L)}}.
	\end{align*}
	Note that $f_1(\omega)$ is different from $f_\delta (\omega)$ in the proof of Proposition 2 because oscillating bandits and monotonic bandits have different expressions of hitting time. Likewise, if $T(\omega ,{\omega ^l},{\omega ^u}) = 2$, define
	\begin{align*}
	{f_2}(\omega ) &\triangleq \left( {L{q^{(K)}} + K{p^{(L)}}} \right)\left( {\frac{{\partial r(\omega ,\lambda )}}{{\partial \lambda }} - \frac{{\partial a(\omega ,\lambda )}}{{\partial \lambda }}} \right)\\
	& = (L - K)\left[ {p - \omega (p + q)} \right](2 - p - q) - 2{q^{(K)}} - 2{p^{(L)}}.
	\end{align*}
	Since $0<2-p-q<1$, it is easy to verify that $f_2(\omega)<0$ if $f_1(\omega)<0$. We can prove that $f_1(\omega^l)<0$ for any $\omega^l\in [0,\omega^*)$ and $f_1(\omega^u)<0$ for any $\omega^u\in (\omega^*,1]$. The proof is given in Lemma C2 in Appendix \hyperref[apd:D]{C}. Therefore, the monotonicity condition holds when $\omega^*\in(\omega^l, \omega^u)$.
	
	Similar to the proof of Proposition 2, we then consider the case that $\omega^*\notin(\omega^l, \omega^u)$. Since $\omega^u\ge \omega^*$, we have three subcases: (i) $\omega^l=\omega^*=\omega^u$; (ii) $\omega^l<\omega^*=\omega^u$; (iii) $\omega^*\le \omega^l <\omega^u$. Specifically,
	\begin{itemize}
		\item[(i)] If $\omega^l=\omega^*=\omega^u$, then $(\omega^l, \omega^u)$ is null; hence the indexability is already proved. 
		\item [(ii)] If $\omega^l<\omega^*=\omega^u$, then $L<\infty$ and $K<\infty$; hence the partial derivatives of $a(\omega,\lambda)$ and $r(\omega,\lambda)$ are the same as given by \eqref{eq:Prop3-da} and \eqref{eq:Prop3-dr1}-\eqref{eq:Prop3-dr2}, respectively. The monotonicity condition still holds and $\omega^u$ will be smaller than $\omega^*$ as $\lambda$ increases. Since $\omega^u\ge \omega^*$ for all $\lambda$,  we know that this case is impossible.
		\item[(iii)] If $\omega^*\le \omega^l <\omega^u$, let us assume there is a $\lambda^*$ such that $\omega^*= \omega^l <\omega^u$, then $L<\infty$ and $K<\infty$. In this case, the optimal average penalty $g=H(\omega^*)$, as stated in Lemma \hyperref[Lem:5]{5}. Meanwhile, recall that		
	\end{itemize}
	\begin{align}  \label{eq:Prop3-g}
	g = \frac{{{q^{(K)}}\left[ {\sum\limits_{k = 1}^L {H\left( {{p^{(k)}}} \right)}  + \lambda } \right] + {p^{(L)}}\left[ {\sum\limits_{k = 1}^K {H\left( {{q^{(k)}}} \right)}  + \lambda } \right]} }{{L{q^{(K)}} + K{p^{(L)}}}}.
	\end{align}
	According to \eqref{eq:Prop3-g}, for a given $\lambda$, $g$ does not change if $K$ and $L$ do not change. Let
	\begin{align*}
	\tilde \omega  \triangleq \min \big\{&\max \{ {p^{(k)}}:{p^{(k)}} < {\omega ^u},k \in {\mathbb{N}^ + }\} ,\\
	&\max \{ 1 - {q^{(k)}}:1 - {q^{(k)}} < {\omega ^u},k \in {\mathbb{N}^ + }\} \big\}. 
	\end{align*}
	Then for any $x\in (\omega^*,\tilde{\omega})$, the threshold policy with sampling region $(x,\omega^u)$ and the threshold policy with sampling region $(\omega^*,\omega^u)$ have the same $L$ and $K$. Therefore, when $\lambda=\lambda^*$, policy $(x,\omega^u)$ and  policy $(\omega^*,\omega^u)$ incur the same average penalty.
	While if $\lambda>\lambda^*$, we must have $\omega^l\ge \tilde{\omega}$; otherwise, $g$ given by \eqref{eq:Prop3-g} will be greater than $H(\omega^*)$.
	In other words, when $\lambda=\lambda^*$ such that $\omega^*= \omega^l <\omega^u$, there are multiple policies, including a unichain policy and some multichain policies, that are equally optimal. Specifically, policy $(\tilde{\omega},\omega^u)$ is an optimal unichain policy; while for any $x\in (\omega^*,\tilde{\omega})$, policy $(x,\omega^u)$ is an optimal multichain policy. Therefore, it is equally optimal to take the active action and the passive action in any $\omega\in [\omega^*,\tilde{\omega}]$ when the service charge is $\lambda^*$; this implies that all belief states in $[\omega^*,\tilde{\omega}]$ share the same Whittle index, i.e., $\lambda^*$.
	
	If $\lambda>\lambda^*$, then $\omega^l\ge \tilde{\omega}$ and the optimal policy is a unichain policy. We may have three cases: (1) $L=\infty, K<\infty$; (2) $L<\infty, K=\infty$; (3) $L=\infty,K=\infty$. It turns out that the monotonicity condition still holds in these cases. See Lemma C3 in Appendix \hyperref[apd:D]{C} for the proof.
	
	In summary, the monotonicity condition holds for any nonempty $(\omega^l, \omega^u)$. We thus proved the indexability of oscillating bandits.
\end{IEEEproof}

\section{Algorithm to Compute Whittle Index}  \label{sec: alg}
To implement the Whittle index policy, we need to compute the Whittle indices for the belief states of each bandit. Since the closed-form expression of Whittle index is unavailable, this section presents an efficient algorithm to compute Whittle index.

The belief state space of a single-bandit is a countably infinite set, i.e., $\Omega  = \{ {p^{(n)}},1 - {q^{(n)}}:n \in {\mathbb{N}^ + }\} $. It is thus intractable (and unnecessary) to compute the Whittle indices of all the belief states. Below, we redefine a finite belief state space by truncating some unnecessary belief states (by unnecessary, we mean that the Whittle indices of the belief states in the neighborhood of $\omega^*$ are approximately equal). Recall that 
\begin{align*}
\mathop {\lim }\limits_{k \to \infty } {p^{(k)}} = \mathop {\lim }\limits_{k \to \infty } \left( {1 - {q^{(k)}}} \right) = {\omega ^ * }.
\end{align*}
For any $\varepsilon>0$, we can find an integer $F<\infty$ such that $|{p^{(F)}} - {\omega ^ * }| < \varepsilon $ and $|1 - {q^{(F)}} - {\omega ^ * }| < \varepsilon $. If $\varepsilon$ is small enough, we can assume that ${p^{(F + 1)}} = 1 - {q^{(F + 1)}} = {\omega ^ * }$ (i.e, we assume the belief states, ${p^{(k)}}, 1 - {q^{(k)}}, k\ge F+1$, have the same Whittle index value as  $\omega^*$). Hence, as far as the Whittle index is concerned,  the belief state space can be viewed as a finite set with $2F+1$ states as follows: 
\begin{align*}
E \triangleq \{ p^{(1)}, \cdots ,{p^{(F)}},{\omega ^ * },1 - {q^{(F)}}, \cdots ,1 - q^{(1)}\}. 
\end{align*}
For concise notation, let $\left| E \right| = 2F + 1$.

Let $({x_1},{x_2}, \cdots ,{x_{|E|}})$ be a permutation of the belief states in $E$ such that $W(x_i)\le W(x_{i+1})$ for all $i$, where $W(x_i)$ is the Whittle index of a belief state $x_i$. In addition, assume that $x_e = \omega^*, 1\le e \le |E|$.
Let $\pi_n$ denote the policy that takes the passive action in ${E_0}(n) \triangleq \{ {x_i}:1 \le i \le n\} $ and the active action in ${E_1}(n) \triangleq \{ {x_i}:n + 1 \le i \le \left| E \right|\} $. Then, $\pi_n$ is the optimal policy of the single-bandit problem with any service charge $\lambda  \in [W({x_n}),W({x_{n + 1}}))$. 
According to Proposition \hyperref[Props:1]{1}, policy $\pi_n$ has a threshold structure. In particular, define
	\begin{align}  \label{eq:VI-1}
	x^l_{n+1} = \min \{\omega:\omega \in E_1(n) \}, \\  \label{eq: VI-2}
	x^u_{n+1} = \max \{\omega:\omega \in E_1(n) \}.
	\end{align}
	 Then $\{\omega:\omega \in E, x^l_{n+1}\le \omega \le x^u_{n+1} \} = E_1(n)$, as shown in Fig. \ref{fig:determinX}. Further, let
	 \begin{align*}
	 \omega^l_{n} = \max \{\omega:\omega \in E_0(n)\cup \{0\}, \omega< x^l_{n+1} \}, \\
	 \omega^u_{n} = \min \{\omega:\omega \in E_0(n)\cup \{1\}, \omega> x^u_{n+1} \}.
	 \end{align*}
Then the hitting time of every $x_i$, under policy $\pi_n$, is given by $T(x_i,\omega^l_n,\omega^u_n)$. For each $\pi_n$, let $L_n \triangleq T(p,\omega^l_n,\omega^u_n)+1$ and $K_n \triangleq T(1-q,\omega^l_n,\omega^u_n)+1$. Note that $L_n$ and $K_n$ can be determined if $\pi_n$ is given. We then have the following proposition:

	 \begin{figure}
	 	\centering
	 	\includegraphics[width=3.5in]{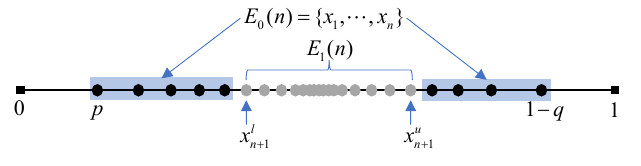}
	 	\caption{Sets $E_0(n)$ and $E_1(n)$. Each dot denotes a point in $E$. Policy $\pi_n$ takes active action in $E_1(n)$.}
	 	\label{fig:determinX}	
	 \end{figure}

\begin{proposition} \label{Props:4}
	For a monotonic bandit ($p+q<1$), the Whittle index of $x_{n+1}$ can be computed as follows: 
	\begin{itemize}
		\item [1. ] For $0\le n\le e-1$, 
		\begin{align*}
		W({x_{n + 1}}) = \frac{{{B_n}H({\tau _{n + 1}}) - {G_n} - \left( {{x_{n + 1}} - {\tau _{n + 1}}} \right){C_n}}}{{\left( {{x_{n + 1}} - {\tau _{n + 1}}} \right)({L_n} - {K_n}) + {q^{({K_n})}} + {p^{({L_n})}}}},
		\end{align*}
		where ${\tau _{n + 1}}\triangleq \tau ({x_{n + 1}})$ and
		\begin{align*}
		&{B_n} = {K_n}{p^{({L_n})}} + {L_n}{q^{({K_n})}},\\
		&{C_n} = {L_n}\sum\limits_{k = 1}^{{K_n}} {H\left( {1-{q^{(k)}}} \right)}  - {K_n}\sum\limits_{k = 1}^{{L_n}} {H\left( {{p^{(k)}}} \right)}, \\
		&{G_n} = {p^{({L_n})}}\sum\limits_{k = 1}^{{K_n}} {H\left( {1-{q^{(k)}}} \right)}  + {q^{({K_n})}}\sum\limits_{k = 1}^{{L_n}} {H\left( {{p^{(k)}}} \right)}. 
		\end{align*}
		\item[2.] For $e\le n <|E|-1$, let $(\omega^l,\omega^u)$ denote the sampling region of the optimal policy with $\lambda = W(x_{n+1})$. If $x_{n+1}=\omega^l$, then $ W(x_{n+1})$ is given by \eqref{eq:Prop4-1}.		  
		 If  $x_{n+1}=\omega^u$, then $ W(x_{n+1})$ is given by  \eqref{eq:Prop4-2}.
		  
		\item[3. ] Finally, $W(x_{|E|})$ can be computed by \eqref{eq:Prop4-1}.
	\end{itemize}
\end{proposition}

\newcounter{Prop4EqCnt} 
\setcounter{Prop4EqCnt}{\value{equation}} 
\setcounter{equation}{52} 
\begin{figure*}[ht] 
	\begin{align} \label{eq:Prop4-1}
	W({x_{n + 1}}) &= \frac{{{q^{({K_n})}} {\sum\limits_{k = 1}^F {\left[ {H\left( {{\tau ^k}({x_{n + 1}})}\right) - {H\left({p^{(k)}}\right) } } \right] } }  - {x_{n + 1}}\sum\limits_{k = 1}^{{K_n}} {\left[ {H\left(1-{q^{(k)}}\right) - H(\omega^*)} \right]} }}{{{q^{({K_n})}} + {x_{n + 1}}}}.	 \\ \label{eq:Prop4-2}
	W({x_{n + 1}}) &= \frac{{{q^{({K_n})}}\left[ {H\left( {{\tau _{n + 1}}} \right) - H(\omega^*)} \right] - \left( {{x_{n + 1}} - {\tau _{n + 1}}} \right)\sum\limits_{k = 1}^{{K_n}} {\left[ {H\left(1-{q^{(k)}}\right) - H(\omega^*)} \right]} }}{{{x_{n + 1}} - {\tau _{n + 1}}}}.
	\end{align}
	\hrulefill  
\end{figure*}

\begin{IEEEproof}
	Statement 1 concerns the case that $\omega^*$ belongs to the sampling region of the optimal policy. Suppose that $({x_1}, \cdots ,{x_{n}})$ has been identified, then the policy $\pi_n$ can be determined. For any $\lambda  \in [W({x_n}),W({x_{n + 1}}))$, $\pi_n$ is the optimal policy. Under policy $\pi_n$, $a({x_{n + 1}},\lambda ) < r({x_{n + 1}},\lambda )$ if $\lambda<W(x_{n+1})$ and $a({x_{n + 1}},\lambda ) = r({x_{n + 1}},\lambda )$ if $\lambda=W(x_{n+1})$. That is,
	\begin{align} \label{eq:Prop4-3}
	a({x_{n + 1}},\lambda ) = {x_{n + 1}}V(1 - q) + \lambda  \le r({x_{n + 1}},\lambda ) = V({\tau _{n + 1}},\lambda ).
	\end{align}
	The above inequality is satisfied with equality if and only if $\lambda=W(x_{n+1})$. Note that the sampling region of policy $\pi_n$ is $\{x_i:n+1\le i\le |E|\}$. Hence $x_e=\omega^*$ belongs to the sampling region of policy $\pi_n$ for any $n<e$. In this case, since $x_{n+1}$ belongs to the sampling region, so does $\tau(x_{n+1})$. Therefore, we have $V({\tau _{n + 1}},\lambda ) = H({\tau _{n + 1}}) - g + a({\tau _{n + 1}},\lambda )$. According to \eqref{eq:Prop4-3}, if $\lambda=W(x_{n+1})$, then 
	\begin{align} \label{eq:Prop4-4}
	({x_{n + 1}} - {\tau _{n + 1}})V(1 - q) = H({\tau _{n + 1}}) - g.
	\end{align}
	As discussed in the proof of Proposition \hyperref[Props:2]{2}, we have
	\begin{align}  \label{eq:Prop4-5}
	&V(1 - q,\lambda ) = \frac{1}{{{q^{({K_n})}}}}\left( {\sum\limits_{k = 1}^{{K_n}} {\left[ {H\left( {1-{q^{(k)}}} \right) - g} \right]}  + \lambda } \right) \\  \label{eq:Prop4-6}
	&g = \frac{{{q^{(K)}}\left[ {\sum\limits_{k = 1}^L {H\left( {{p^{(k)}}} \right)}  + \lambda } \right] + {p^{(L)}}\left[ {\sum\limits_{k = 1}^K {H\left( {1-{q^{(k)}}} \right)}  + \lambda } \right]} }{{L{q^{(K)}} + K{p^{(L)}}}}. 
	\end{align}
	Solving \eqref{eq:Prop4-4}-\eqref{eq:Prop4-6} yields statement 1 in Proposition \hyperref[Props:4]{4}.
	
	Statement 2 concerns the case that $\lambda>W(x_e)$, i.e., $\omega^*$ does not belong to the sampling region of the optimal policy. In particular, for $n>e$ and $\lambda  \in [W({x_n}),W({x_{n + 1}}))$,  we have $\{ {x_{n + 1}}, \cdots ,{x_{|E|}}\}  \subseteq ({\omega ^l},{\omega ^u})$, where ${\omega ^ * } \le {\omega ^l} < {\omega ^u}$. Whittle index $W(x_{n+1})$ equals to the $\lambda$ such that $x_{n+1}$ is either $\omega^l$ or $\omega^u$. Also, we have ${x_{n + 1}}V(1 - q) + \lambda  = V({\tau _{n + 1}},\lambda )$. According to the Bellman equation,
	\begin{align}  \label{eq:Prop4-7}
	&{q^{({K_n})}}V(1 - q) = \sum\limits_{j = 1}^{{K_n}} {\left[ {H\left( {1 - {q^{(j)}}} \right) - g} \right]} +  \lambda, \\  \label{eq:Prop4-8}
	&V(p) = \sum\limits_{j = 1}^F {\left[ {H\left( {{p^{(j)}}} \right) - g} \right]}  + V({\omega ^ * }) = 0,
	\end{align}
	where $g=H(\omega^*)$. In \eqref{eq:Prop4-8}, we use $F$ instead of $\infty$ because we assume $F$ is large enough such that $p^{(F)}=\omega^*$. If $x_{n+1}=\omega^l$, then ${\tau ^k}({x_{n + 1}}) \notin ({\omega ^l},{\omega ^u})$ for all $k$. Hence
	 \begin{align} \label{eq:Prop4-9}
	 V\left( {{\tau _{n + 1}}} \right) = \sum\limits_{j = 1}^F {\left[ {H\left( {{\tau ^j}({x_{n + 1}})} \right) - g} \right]}  + V({\omega ^ * }).
	 \end{align}
	 Solving the above equations gives \eqref{eq:Prop4-1} in Proposition \hyperref[Props:4]{4}.
	 
	 On the other hand, if $x_{n+1}=\omega^u$ and $n<|E|-1$, then ${\tau }({x_{n + 1}}) \in ({\omega ^l},{\omega ^u})$. Hence
	 \begin{align} \label{eq:Prop4-10}
	 V\left( {{\tau _{n + 1}}} \right) = H\left( {{\tau _{n + 1}}} \right) - g + {\tau _{n + 1}}V(1 - q) + \lambda. 
	 \end{align}
	 From \eqref{eq:Prop4-7},\eqref{eq:Prop4-8} and \eqref{eq:Prop4-10}, we can obtain \eqref{eq:Prop4-2}.
	 
	 Finally, when $\lambda=W(x_{|E|})$, $V(\tau(x_{|E|}))$ is of the form given by \eqref{eq:Prop4-9} because $(\omega^l,\omega^u)$ does not contain any belief state in $E$. Hence $W(x_{|E|})$ can be computed using \eqref{eq:Prop4-1}.
\end{IEEEproof}

Proposition 4 implies an iterative algorithm to identify the permutation $({x_1},{x_2}, \cdots ,{x_{|E|}})$  and the corresponding Whittle indices. Recall that the sampling region monotonically shrinks as $\lambda$  increases. Given $({x_1}, \cdots ,{x_{n}})$,  $x_{n+1}$ can be either $x_{n+1}^l$ or $x_{n+1}^u$, see Fig. \ref{fig:determinX}. Note that \eqref{eq:VI-1} and \eqref{eq: VI-2} are equivalent to
\begin{align} \label{eq:xi-l}
x_{n+1}^l = \min \{\omega:\omega\in E,\omega\notin \{{x_1}, \cdots ,{x_{n}}\} \},  \\ \label{eq:xi-u}
x_{n+1}^u = \max \{\omega:\omega\in E,\omega\notin \{{x_1}, \cdots ,{x_{n}}\} \}.
\end{align}
As $\lambda$ increases from $W(x_n)$ to $W(x_{n+1})$, $\omega^l$ moves toward $x_{n+1}^l$, while $\omega^u$ moves toward $x_{n+1}^u$. 
We then compute the Whittle index for each of the two options by Proposition \hyperref[Props:4]{4}, and the option with the smaller index value is $x_{n+1}$. Note that if $\lambda<W(x_1)$, the optimal policy $\pi_0$ takes the active action in all belief states; hence $L_0=1$ and $K_0=1$. Start from $\pi_0$, we can identify each pair of belief state and Whittle index, $\big({x_1},W({x_1})\big) \cdots \big({x_{|E|}},W({x_{|E|}})\big)$. The details of the procedure are presented in Algorithm \ref{al:CWI}.

\begin{algorithm}[!ht]
	\caption{Computing the Whittle index ($p+q<1$) }
	\label{al:CWI}
	\KwIn{Transition probabilities $p,q$; cut-off factor $F$. \\		
	}
	\textbf{Initialize:} $ L_0\leftarrow1, K_0\leftarrow1$, $ i\leftarrow0 $.  \\
	\qquad \qquad $\omega^* \leftarrow p/(p+q)$. \\
	\qquad \qquad $E\leftarrow\{\omega^*, p^{(k)},1-q^{(k)}:k=1,\cdots,F\}$.  \\
	\qquad \qquad $\{x_i=0:i=1,2,\cdots,(2F+1)\}$.  \\
	\While{$L_i\le F+1$}{
	$i= i+1$.  \\
	Determine $x_i^l$ by \eqref{eq:xi-l} and $x_i^u$ by \eqref{eq:xi-u}.  \\
	Compute $W(x_i^l)$ and $W(x_i^u)$ by statement 1 of Proposition \hyperref[Props:4]{4}.  \\
	$x_i=\arg \min \{W(x_i^l),W(x_i^u)\}$.  \\
	$W(x_i) = \min \{W(x_i^l),W(x_i^u)\}$. \\
	Determine $L_i$ and $K_i$ based on $(x_1,\cdots,x_i)$.
	}
	
	\While{$i\le 2F$}{
		$i= i+1$. \\
		Determine $x_i^l$ by \eqref{eq:xi-l} and $x_i^u$ by \eqref{eq:xi-u}.  \\
		Compute $W(x_i^l)$ by \eqref{eq:Prop4-1} and $W(x_i^u)$ by \eqref{eq:Prop4-2}. \\
		$x_i=\arg \min \{W(x_i^l),W(x_i^u)\}$.  \\
		$W(x_i) = \min \{W(x_i^l),W(x_i^u)\}$. \\
		Determine $K_i$ based on $(x_1,\cdots,x_i)$.
	}
	$x_{|E|}\leftarrow E-\{x_1,\cdots,x_{2F}\}$.  // set difference  \\
	Compute $W(x_{|E|})$ by \eqref{eq:Prop4-1}.		
	
\end{algorithm}	
  
The approach to computing Whittle index for oscillating bandits is a bit different from monotonic bandits. Again, the difference is caused by the oscillating feature of $\tau^k(\omega)$. We have the following proposition for the Whittle index of an oscillating bandit.

\newcounter{Prop5EqCnt} 
\setcounter{Prop5EqCnt}{\value{equation}} 
\setcounter{equation}{64} 
\begin{figure*}[ht] 
	\begin{align} \label{eq:Prop5-1}
	W({x_{n + 1}}) = 
	\frac{{{B_n}\left[ {H({\tau _{n + 1}}) + H(\tau _{n + 1}^{(2)})} \right] - 2{G_n} - \left( {{x_{n + 1}} - \tau _{n + 1}^{(2)}} \right){C_n}}}{{\left( {{x_{n + 1}} - \tau _{n + 1}^{(2)}} \right)({L_n} - {K_n}) + 2\left( {{q^{({K_n})}} + {p^{({L_n})}}} \right)}}. 	
	\end{align}
	\hrulefill  
\end{figure*}

\begin{proposition}  \label{Props:5}
	For an oscillating bandit ($p+q>1$), the Whittle index of $x_{n+1}$ can be computed as follows:
	\begin{itemize}
		\item [1.] For $0\le n\le e$. Let ${\tau _{n + 1}} \triangleq \tau ({x_{n + 1}}),\tau _{n + 1}^{(2)} \triangleq {\tau ^2}({x_{n + 1}})$. If $\tau_{n + 1} \in \{x_1,\cdots,x_n\}$, $W(x_{n+1})$ is given by \eqref{eq:Prop5-1}.
		If $\tau_{n + 1} \notin \{x_1,\cdots,x_n\}$, then
		\begin{align*}
		W({x_{n + 1}}) = \frac{{{B_n}H({\tau _{n + 1}}) - {G_n} - \left( {{x_{n + 1}} - {\tau _{n + 1}}} \right){C_n}}}{{\left( {{x_{n + 1}} - {\tau _{n + 1}}} \right)({L_n} - {K_n}) + {q^{({K_n})}} + {p^{({L_n})}}}},
		\end{align*}	
		where $B_n, C_n$ and $G_n$ are defined in Proposition \hyperref[Props:4]{4}.				 

	 	\item [2. ] For $e\le n <|E|$, $W(x_n)=W(x_e)=W(\omega^*)$.
	 	\item [3. ] Finally, $W(x_{|E|})$ can be computed by \eqref{eq:Prop4-1}.	 	
	 \end{itemize}
\end{proposition}  
\begin{IEEEproof}
	Statement 1 can be proved by a similar argument as in the proof of Proposition \hyperref[Props:4]{4}. However, since $\tau^k(x_{n+1})$ is an oscillating function of $k$, $\tau(x_{n+1})$ may not belong to the sampling region $(\omega^l,\omega^u)$ when $x_{n+1}$ equals to $\omega^l$ or $\omega^u$. Under policy $\pi_n$, if $\tau(x_{n+1})\in(\omega^l,\omega^u)$, i.e., $\tau(x_{n+1})\notin \{x_1,\cdots,x_n\}$, then 
	\begin{align} \label{eq:Prop5-2}
	V({\tau _{n + 1}},\lambda ) = H({\tau _{n + 1}}) - g + a({\tau _{n + 1}},\lambda ).
	\end{align}
	On the other hand, if $\tau(x_{n+1})\notin(\omega^l,\omega^u)$, i.e., $\tau(x_{n+1})\in \{x_1,\cdots,x_n\}$, then 
	\begin{align} \label{eq:Prop5-3}
	V({\tau _{n + 1}},\lambda ) = H({\tau _{n + 1}}) + H(\tau _{n + 1}^{(2)}) - 2g + a(\tau _{n + 1}^{(2)},\lambda ).
	\end{align}
	Substituting $V(1-q,\lambda)$ given by \eqref{eq:Prop4-5} and $g$ given by \eqref{eq:Prop4-6} into \eqref{eq:Prop5-2} and \eqref{eq:Prop5-3}, respectively, we can obtain statement 1.
	
	Statement 2 is an immediate result from Proposition \hyperref[Props:3]{3}. Finally, if  $\lambda  = W({x_{|E|}})$, the sampling region does not contain any state in $E$. Then statement 3 can be proved by the same argument as in Proposition \hyperref[Props:4]{4}.
\end{IEEEproof}

With Proposition 5, we can apply a similar method as  Algorithm 1 to compute the Whittle index for oscillating bandits. Therefore, for any bandit with $p,q\in [0,1]$  ($p+q\neq 0,1$ and $2$), the Whittle index can be computed. Hence we can implement the Whittle index policy for the UoI scheduling problem.

\section{Discussion} \label{sec:discussion}
In this section, we first study a special case in which UoI reduces to a nonlinear function of AoI to build a connection between UoI and AoI. Then, we explain how the results in this paper apply not only to the UoI scheduling problem but also to a class of RMABs with a concave penalty function of the belief state. We also explore possible directions for future work.

\subsection{A Special Case Study}
This part examines the special case of the single-bandit problem in which $p=q$. The transition probabilities of the associated Markov process satisfy $P[0|1]=P[1|0]$, hence UoI is not affected by the value of the last observation and is a function of AoI only. Consequently, equations used to compute the Whittle index in Propositions \hyperref[Props:4]{4} and \hyperref[Props:5]{5} reduce to a closed-form expression. This expression is consistent with a previous study on AoI-based scheduling \cite{AoIcost_nonlinear_MIT}.

Before we derive the expression of the Whittle index for the special case, we need the following result.

\begin{lemma}
	For a single-bandit problem with $\lambda\ge 0$ and $p=q\in (0,1)$, we have
	\begin{itemize}
		\item[1.] $V(\omega) = V(1-\omega),\forall \omega \in (0,1)$
		\item[2.] The optimal sampling region $(\omega^l,\omega^u)$ is symmetric about the point $0.5$, i.e., $\omega^u = 1-\omega^l$.
	\end{itemize}
\end{lemma}
\begin{IEEEproof}
	Recall that the belief state $\omega$ is defined as the probability that the underlying Markov process is in state 1. That is, at any time  $t$, the belief state being  $\omega$ means $P\{S(t)=1\}=\omega$, where $S(t)\in \{0,1\}$  is the state of the remote Markov process. There is an alternative way to define the belief state. Specifically, let $\eta = 1-\omega$ denote the probability that the underlying Markov process is in state 0. In the following, we call $\omega$  the type-1 belief state and $\eta$  the type-0 belief state. For the same single-bandit problem, the Bellman equation w.r.t $\eta$  is
	\begin{align} \label{eq:Lem9-1}
	&H(\eta ) + \min \{ \lambda  + \eta \hat V(1 - p) + (1 - \eta )\hat V(p),\hat V(\hat \tau (\eta ))\} \nonumber \\
	=&\hat V(\eta ) + g, 
	\end{align} 
	where $\hat{\tau}(\eta)\triangleq \eta (1-p)+(1-\eta)p$. Since $p=q$, we have $\hat{\tau}(\eta) = \eta(1-2p)+p = \tau(\eta)$. Replacing $\hat{\tau}$ with $\tau$ in \eqref{eq:Lem9-1} yields
	\begin{align}\label{eq:bellman-eta}
	&H(\eta ) + \min \{ \lambda  + \eta \hat V(1 - p) + (1 - \eta )\hat V(p),\hat V(\tau (\eta ))\} \nonumber \\
	=&\hat V(\eta ) + g.
	\end{align}
	Note that \eqref{eq:bellman-eta} and \eqref{eq:bellmaneq} have the same form. Since the Bellman equation has a unique solution up to a constant, if we set $\hat{V}(p)=V(p)=0$, then we must have $\hat{V}(x)=V(x)$ for $x\in (0,1)$. However, the physical meaning of $\hat{V}(x)$ and $V(x)$ are different. The value $V(x)$ is defined based on type-1 belief state, which can be expressed as \cite{puterman1994markov}:
	\begin{align}
	V(x) = E\left[ {\sum\limits_{t = 1}^\infty  (U[t] - g) |P\{ S(1) = 1\}  = x} \right],
	\end{align}
	where $U[t]$ denotes the UoI at time $t$. On the other hand, $\hat{V}(x)$ is defined based on type-0 belief state $\eta = 1-\omega$, therefore,
	\begin{align*}
	\hat V(x)& = E\left[ {\sum\limits_{t = 1}^\infty  (U[t] - g) |P\{ S(1) = 0\}  = x} \right] \nonumber \\
	&= E\left[ {\sum\limits_{t = 1}^\infty  (U[t] - g) |P\{ S(1) = 1\}  = 1 - x} \right] = V(1 - x).
	\end{align*}
	Since $\hat{V}(x)=V(x)$ for all $x\in (0,1)$, we conclude that $V(\omega)=V(1-\omega)$.
	
	For statement 2, note that $V(1-p)=V(p)=0$, hence $a(\omega)=\lambda$ is a constant. Also, $r(\omega)=V(\tau)=V(1-\tau)=r(1-\omega)$. Clearly, if $a(x)=r(x)$, then $a(1-x)=r(1-x)$. We thus conclude that $\omega^u = 1-\omega^l$.		
	
\end{IEEEproof}

Lemma 9 states that both the value function and the optimal sampling region are symmetric about $\omega=0.5$. Note that if $p=q$, then $\omega^*=0.5$. Hence $\omega^* \in (\omega^l,\omega^u)$ whenever $(\omega^l,\omega^u)$ is non-empty. As far as the optimal action is concerned, it does not matter whether we are in belief state $p^{(n)}$ or belief state $1-p^{(n)}$. This verifies our intuition that ,if $p=q$, UoI is independent of the value of the last observation and only depends on AoI. The Whittle index of this case has a closed-form expression, as given below. 
\begin{corollary}
	For a bandit with symmetric transition probabilities $p=q\in(0,1)$, the Whittle indices of belief states $p^{(n)}$ and $1-p^{(n)}$ are given by
	\begin{align*}
	W({p^{(n)}}) = W(1 - {p^{(n)}}) = \sum\limits_{k = 1}^n {\left[ {H({p^{(n + 1)}}) - H({p^{(k)}})} \right]}, 
	\end{align*}
	where $n=1,2,3,\cdots$.
\end{corollary}
\begin{IEEEproof}
	We first consider the case of $p=q<0.5$. As discussed above, $\omega^* \in (\omega^l,\omega^u)$ whenever $(\omega^l,\omega^u)$ is non-empty. We thus apply statement 1 in Proposition \hyperref[Props:4]{4} to compute Whittle indices of all belief states. Since $\omega^u = 1-\omega^l$ and $\omega^* = 0.5$, it is easy to verify from Lemma \hyperref[Lem:7]{7} that $T(p,\omega^l,\omega^u) = T(1-p,\omega^l,\omega^u)$. Therefore, we have $L_n=K_n$ in Proposition \hyperref[Props:4]{4}, for all $n$. The expression in Corollary 2 can be obtained immediately by substituting $L_n = K_n$ and $p=q$ into statement 1 of Proposition 4. The case of $p=q>0.5$ can be proved similarly from Lemma \hyperref[Lem:8]{8} and Proposition \hyperref[Props:5]{5}.
\end{IEEEproof}

Note that the parameter $n$ in Corollary 2 is the AoI of the last observation. Corollary 2 is consistent with the result in \cite{AoIcost_nonlinear_MIT}. In \cite{AoIcost_nonlinear_MIT}, a non-decreasing penalty function of AoI is introduced as a metric of information freshness. In the case of reliable channels, the Whittle index obtained in \cite{AoIcost_nonlinear_MIT} is identical to ours in Corollary 2.

As mentioned, if $p=q$, UoI reduces to a nonlinear and non-decreasing function of AoI, hence the AoI-based methods may be applied to the problem of UoI-based scheduling. However, in the general case where $p\neq q$, given different values of the last observation, the quality of information evolves with time at different rates, as shown by the examples in Section II. In general, AoI does not evolve in a way that depends on the last observation, while UoI does because its evolution depends on the last observed belief state.

\subsection{Extension and Discussion}
We have developed a Whittle index policy for the problem of UoI scheduling. In fact, our method applies not only to the UoI problem but also to a class of RMABs with a concave penalty function of the belief state. Formally, we define a class of RMABs as follows:

\begin{definition}[C-type RMAB]
	An RMAB is called C-type if it is of the form of problem P1 and the penalty function $H(\omega)$ is a concave function of the belief state $\omega$. 
\end{definition}

Clearly, the UoI-scheduling problem is a C-type RMAB. More examples can be found in our simulation; an RMAB with any penalty function in Section \ref{sec:simul} is C-type. We next show that the Whittle index policy developed for the UoI-scheduling problem can be directly extended to any C-type RMAB.

\begin{theorem}
	A C-type RMAB is indexable. The Whittle indices of monotonic bandits and oscillating bandits can be computed by Proposition \hyperref[Props:4]{4} and Proposition \hyperref[Props:5]{5}, respectively.
\end{theorem}
\begin{IEEEproof}
	This theorem follows directly from the analysis in previous sections. Just note that Propositions 1-5 are all valid as long as the function $H(\omega)$ is concave w.r.t. $\omega$.
\end{IEEEproof}

The UoI metric adopted in this paper requires knowledge of the model, i.e., the transition probabilities of the remote Markov processes. To apply the UoI metric in practical systems, we need to learn the model first. If the system is stationary, the transition probabilities can be learned by proper sampling algorithms given enough observations. How to learn the model efficiently is an interesting problem for future work.

This paper studies the case of binary Markov processes only. In a more general setting, the remote processes may have multiple ($\ge 2$) states. It would be interesting to extend our results to a C-type RMAB with multi-state Markov processes. Establishing indexability for this general model is challenging. The belief state space for a $k$-state Markov process is a $(k-1)$-simplex. When  $k>2$, the belief state is no longer a scalar but a vector; consequently, the value function must be defined on a multi-dimensional space. This significantly increases the difficulty of proving indexability, although we conjecture that the problem is still indexable. Doing so awaits future study.

\section{Simulations}  \label{sec:simul}
This section presents numerical results that demonstrate the excellent performance of the Whittle index policy. We first present the performance of the Whittle index policy for the UoI scheduling problem. Then, we consider general RMABs with different concave penalties and show that the Whittle index policy also performs well in these cases. Throughout this section, the optimal policy is found using relative value iteration. The results of a myopic policy are also presented for benchmarking purposes. In particular, at the beginning of each time slot $t$, the myopic policy computes the one-step penalty $H(\omega_i(t))$ for each process $i$ and selects the $m$ processes with the largest $m$ one-step penalties to update in this time slot. We first present some results for the case of single channel (i.e., only one process is selected at each time). Simulations for the case of multiple channels ($m>1$) are provided in the last table.

Table \ref{tab:UoI-1} compares the time-averaged sum-UoIs of the three policies in systems with 2 processes. For each setting, we conducted 50 independent runs, with each run lasting $10^4$ slots. We compute the time-averaged sum-UoIs for each run. Then the performance of each policy is evaluated by averaging the results over the 50 independent runs.
The regret of the Whittle index policy is computed by $(g_w - g^*)/g^*$, where $g_w$ and $g^*$ denote the average sum-UoIs of the Whittle index policy and the optimal policy, respectively. The regret of the myopic policy is computed similarly. As shown in Table \ref{tab:UoI-1}, the Whittle index policy obtains near-optimal performance in all settings (regrets within $0.1\%$). By contrast, the myopic policy has significant regrets in some settings. 
\begin{table}[h]
	\centering
	\begin{tabular}{|l|l|l|l|l|l|l|}
		\hline
		& $(p_i,q_i)$                                                          & \begin{tabular}[c]{@{}l@{}}Optimal\\ UoI\end{tabular} & \begin{tabular}[c]{@{}l@{}}WI\\ UoI\end{tabular} & \begin{tabular}[c]{@{}l@{}}Myopic\\ UoI\end{tabular} & \begin{tabular}[c]{@{}l@{}}WI\\ regret\end{tabular} & \begin{tabular}[c]{@{}l@{}}Myopic\\ regret\end{tabular} \\ \hline
		A1 & \begin{tabular}[c]{@{}l@{}}(0.05,0.2)\\ (0.2,0.4)\end{tabular} 
		& 1.2866  & 1.2867  & 1.527   & 0.01\%    & 18.7\%    \\ \hline
		A2 & \begin{tabular}[c]{@{}l@{}}(0.2,0.2)\\(0.4,0.4)\end{tabular}                                                              &1.7219 & 1.7219  &  1.873 & 0 &8.8\%    \\ \hline
		A3 & \begin{tabular}[c]{@{}l@{}}(0.95,0.95)\\(0.7,0.7)\end{tabular}                                                               & 1.2864 & 1.2864 & 1.5668 & 0 & 21.8\% \\ \hline
		A4 & \begin{tabular}[c]{@{}l@{}}(0.05,0.1)\\(0.2,0.9)\end{tabular}                                                                 & 1.0309 & 1.0318 & 1.2424 & 0.09\% & 20.5\%  \\ \hline
	\end{tabular}
	\caption{Time-averaged sum-UoIs of the optimal policy, the Whittle index policy (WI), and the myopic policy in 2-process systems, wherein $(p_i,q_i)$ denotes the state-transition probabilities of the process $i$.}
	\label{tab:UoI-1}
\end{table}

Observe that the Whittle index policy attains the exact optimality in A2 and A3. In the two settings, the remote Markov processes are symmetric (i.e., $p_i=q_i$). For the case of $p_i = q_i \in (0,0.5)$, UoI is a monotonically increasing function of age; this special case falls into the scope of \cite{AoIcost_nonlinear_MIT}, wherein the authors proved that the Whittle index policy is optimal in 2-process systems.

We next evaluate the performance of the Whittle index policy for the UoI scheduling problem in 3-process systems. We consider three settings with different transition probabilities. As shown in Table \ref{tab:UoI-2}, the Whittle index policy also exhibits near-optimal performance in 3-process systems. We remark that the complexity of computing the optimal policy by relative value iteration increases exponentially as the number of processes (i.e., $M$) increases. For the UoI scheduling problem, it becomes quite difficult for the relative value iteration to converge in systems with more than 3 processes. The Whittle index policy, however, is easy to implement.  The complexity of computing the Whittle index by our algorithm increases linearly with $M$.

\begin{table}[h]
	\centering
	\begin{tabular}{|l|l|l|l|l|l|l|}
		\hline
		& $(p_i,q_i)$                                                          & \begin{tabular}[c]{@{}l@{}}Optimal\\ UoI\end{tabular} & \begin{tabular}[c]{@{}l@{}}WI\\ UoI\end{tabular} & \begin{tabular}[c]{@{}l@{}}Myopic\\ UoI\end{tabular} & \begin{tabular}[c]{@{}l@{}}WI\\ regret\end{tabular} & \begin{tabular}[c]{@{}l@{}}Myopic\\ regret\end{tabular} \\ \hline
		B1 & \begin{tabular}[c]{@{}l@{}}(0.1,0.1)\\ (0.6,0.6)\\ (0.3,0.3) \end{tabular} 
		& 2.469  & 2.469  & 2.792  & 0    & 13.8\%    \\ \hline
		B2 &\begin{tabular}[c]{@{}l@{}}(0.1,0.3)\\(0.6,0.6)\\ (0.1,0.2)\end{tabular}                                       
		&2.2963 & 2.2968  &  2.7005 & 0.02\% &17.6\%    \\ \hline
		B3 & \begin{tabular}[c]{@{}l@{}}(0.1,0.3)\\(0.5,0.6)\\ (0.9,0.9)\end{tabular}                                                             
		& 2.2158 & 2.2179 & 2.6506 & 0.1\% & 19.6\% \\ \hline		
	\end{tabular}
	\caption{Time-averaged sum-UoIs of the optimal policy, the Whittle index policy (WI), and the myopic policy in 3-process systems. }
	\label{tab:UoI-2}
\end{table}

\begin{table}[h]
	\centering
	\begin{tabular}{|l|l|l|l|l|l|l|}
		\hline
		& $(p_i,q_i)$                                                          & \begin{tabular}[c]{@{}l@{}}Optimal\\ penalty\end{tabular} & \begin{tabular}[c]{@{}l@{}}WI\\ penalty\end{tabular} & \begin{tabular}[c]{@{}l@{}}Myopic\\ penalty\end{tabular} & \begin{tabular}[c]{@{}l@{}}WI\\ regret\end{tabular} & \begin{tabular}[c]{@{}l@{}}Myopic\\ regret\end{tabular} \\ \hline
		C1 & \begin{tabular}[c]{@{}l@{}}(0.05,0.2)\\ (0.4,0.5)\end{tabular} 
		& 1.057  & 1.064  & 1.275   & 0.7\%    & 20.6\%    \\ \hline
		C2 & \begin{tabular}[c]{@{}l@{}}(0.05,0.1)\\ (0.5,0.6) \end{tabular}                                                              
		&1.480 & 1.482  &  1.814 & 0.1\% &22.6\%    \\ \hline
		D1 & \begin{tabular}[c]{@{}l@{}}(0.05,0.2)\\(0.1,0.3)\\(0.4,0.7)\end{tabular}                                                               & 1.1467 & 1.1485 & 1.4079 & 0.2\% & 22.8\% \\ \hline
		D2 & \begin{tabular}[c]{@{}l@{}}(0.1,0.2)\\(0.1,0.8)\\ (0.4,0.5)\end{tabular}                                                                 
		& 1.3843 & 1.3845 & 1.587 & 0.01\% & 14.6\%  \\ \hline
	\end{tabular}
	\caption{The RMAB with penalty function $H_1(\omega)$, where $\alpha_0=-1,\alpha_1=2,\beta=0.5$.  C1 and C2 are 2-process systems. D1 and D2 are 3-process systems with one channel.}
	\label{tab:penaltyH1}
\end{table}
\begin{table}[b]
	\centering
	\begin{tabular}{|l|l|l|l|l|l|l|}
		\hline
		& $(p_i,q_i)$                                                          & \begin{tabular}[c]{@{}l@{}}Optimal\\ penalty\end{tabular} & \begin{tabular}[c]{@{}l@{}}WI\\ penalty\end{tabular} & \begin{tabular}[c]{@{}l@{}}Myopic\\ penalty\end{tabular} & \begin{tabular}[c]{@{}l@{}}WI\\ regret\end{tabular} & \begin{tabular}[c]{@{}l@{}}Myopic\\ regret\end{tabular} \\ \hline
		E1 & \begin{tabular}[c]{@{}l@{}}(0.05,0.2)\\ (0.4,0.5)\end{tabular} 
		& 1.2677  & 1.268  & 1.618   & 0.02\%    & 27.6\%    \\ \hline
		E2 & \begin{tabular}[c]{@{}l@{}}(0.05,0.2)\\ (0.4,0.5)\\ (0.1,0.2) \end{tabular}                                                              
		&1.904 & 1.906  &  2.507 & 0.1\% &31.7\%    \\ \hline
		F1 & \begin{tabular}[c]{@{}l@{}}(0.05,0.2)\\(0.4,0.5)\end{tabular}                                                               & 21.466 & 21.622 & 32.722 & 0.7\% & 52.4\% \\ \hline
		F2 & \begin{tabular}[c]{@{}l@{}}(0.05,0.2)\\ (0.4,0.5)\\ (0.1,0.2)\end{tabular}                                                                 
		& 37.875 & 38.225 & 49.722 & 0.9\% & 32.3\%  \\ \hline
	\end{tabular}
	\caption{The RMABs with different penalties. For E1 and E2, the penalty is $H_2(\omega)=1-(2\omega-1)^2$. For F1 and F2, the penalty is $H_3(\omega)=20-1/\omega$. }
	\label{tab:others}
\end{table}

The Whittle index policy developed in this paper applies to a class of RMABs, wherein the UoI scheduling problem is only a special case.  To show this, we apply the Whittle index policy to RMABs with different penalties than UoI.  First, we construct a penalty function with the following form:
\begin{align*}
{H_1}(\omega ) &= {\alpha _1}\omega  + {\alpha _0}(1 - \omega )  \\
&+ \beta \sqrt {\alpha _1^2\omega  + \alpha _0^2(1 - \omega ) - {{\left( {{\alpha _1}\omega  + {\alpha _0}(1 - \omega )} \right)}^2}}. 
\end{align*}
The meaning of ${H_1}(\omega )$  can be interpreted as follows: suppose that a cost $\alpha_i$ is generated in each time slot if the underlying Markov process is in state $i,i\in \{0,1\}$. Then ${\alpha _1}\omega  + {\alpha _0}(1 - \omega )$ is the expected one-step cost, and the last term of ${H_1}(\omega )$ is the standard deviation with a weight coefficient $\beta > 0$. Therefore, using $H_1(\omega)$ as the penalty means minimizing a weighted linear combination of the expected cost and the standard deviation of the cost. Note that $H_1(\omega)$ is concave w.r.t. $\omega$. In Table \ref{tab:penaltyH1}, we choose $\alpha_0=-1,\alpha_1=2,\beta=0.5$. The results show that the Whittle index policy also achieves near-optimal performance---regrets within $1\%$---for the RMAB with penalty function $H_1(\omega)$.

Table \ref{tab:others} presents the results of RMABs with different penalties than UoI and $H_1(\omega)$. In E1 and E2, the penalty is $H_2(\omega)=1-(2\omega-1)^2$. In F1 and F2, the penalty is $H_3(\omega)=20-1/\omega$. The two penalties are concave functions of the belief state. In addition, E1 and F1 are 2-process systems, while E2 and F2 are 3-process systems with one channel. As shown in the table, the Whittle index policy has tiny regrets (smaller than $1\%$) in all settings.  The results demonstrate that the Whittle index policy developed in this paper applies to the general class of RMABs wherein the penalty is a concave function of the belief state. Although the Whittle index policy can not achieve exact optimality in general, the near-optimal performance and low complexity make it an excellent algorithm for these RMABs.

Finally, Table \ref{tab:multiple} provides some simulation results for the case of multiple channels. It is difficult to compute the optimal policies for systems in Table  \ref{tab:multiple} because of the associated exponential computation complexity. In place of the optimal policy, we introduce the round-robin (RR) scheme for an additional benchmark here. It was proved in \cite{Kadota2018TON} that the round-robin scheme is the optimal policy to minimize the average AoI in the case of reliable transmissions; hence the round-robin scheme can also be viewed as an AoI-based policy. The results in Table \ref{tab:multiple} show again that the Whittle index policy outperforms the other two methods. In addition, we note that the round-robin scheme is better than Myopic policy in most cases. The reason perhaps is related to the positive relationship between UoI (and other concave functions of the belief state) and AoI, as we discussed under Corollary \hyperref[Coro:1]{1}. From this perspective, the AoI-optimal policy can be viewed as a greedy policy for our problem. Hence it is not surprising that the AoI-optimal policy (i.e., RR) is better than the Myopic policy.
\begin{table*}[t]
	\centering
	\begin{tabular}{|l|l|l|l|l|l|l|l|}
		\hline
		& Penalty & $(M,m)$                                                           & \begin{tabular}[c]{@{}l@{}}WI\\ penalty\end{tabular} & \begin{tabular}[c]{@{}l@{}}Myopic\\ penalty\end{tabular} & \begin{tabular}[c]{@{}l@{}}RR\\ penalty\end{tabular} & \begin{tabular}[c]{@{}l@{}}WI vs.\\ Myopic\end{tabular} & \begin{tabular}[c]{@{}l@{}}WI vs.\\ RR\end{tabular} \\ \hline
		G1& UoI & \begin{tabular}[c]{@{}l@{}}(5,2)\end{tabular} 
		& 3.53 & 4.64  & 4.06   & 31.6\%    & 15\%    \\ \hline
		G2 & UoI & \begin{tabular}[c]{@{}l@{}}(10,3) \end{tabular}                                                              
		&5.14 & 5.71  &  5.64 & 11\% &9.7\%    \\ \hline
		G3 & $H_1(\omega)$ & \begin{tabular}[c]{@{}l@{}}(10,3)\end{tabular}                                                               & 1.15 & 1.35 & 1.38 & 17.2\% & 19.9\% \\ \hline
		G4 & $H_2(\omega)$ & \begin{tabular}[c]{@{}l@{}}(10,3)\end{tabular}                                                                 
		& 5.47 & 5.96 & 5.93 & 8.9\% & 8.6\%  \\ \hline
		G5 & $H_3(\omega)$ & \begin{tabular}[c]{@{}l@{}}(10,3)\end{tabular}                                                                 
		& 63.67 & 81.06 & 75.2 & 27.3\% & 18.1\%  \\ \hline
	\end{tabular}
	\caption{The RMABs with different penalties in multi-channel systems. $M$ is the number of processes, $m$ is the number of channels. Penalties $H_i(\omega),i=1,2,3$, are defined above. WI vs. Myopic is computed by $(g_M-g_w)/g_w$, where $g_M$ and $g_w$ denote the average penalties of Myopic and WI, respectively. WI vs. RR is computed similarly. }
	\label{tab:multiple}
\end{table*}

\section{Conclusion}   \label{sec:con}
This paper adopted UoI as a metric for quantifying information freshness. We considered a system in which a central monitor observes multiple binary Markov processes through $m$ communication channels. At each time step, $m$ of the Markov processes are scheduled to update their state information to the monitor. The UoI of a Markov process corresponds to the monitor's uncertainty about its state. We studied the scheduling policies that minimize the time-averaged sum-UoI of the processes. We formulated the UoI scheduling problem as an RMAB. In this formulation, each bandit of the RMAB is a POMDP, and its UoI is a concave function of its belief state. We developed a Whittle index scheduling policy for the RMAB through the following steps:
\begin{itemize}
	\item [1)] We first analyzed the single-bandit problem associated with the RMAB and established the threshold structure of the single-bandit problem’s optimal policy. 
	\item [2)] We applied the threshold structure to develop a sufficient condition for the indexability of a single bandit, i.e., the monotonicity condition. In addition, we derived the hitting time for a belief state to evolve to the sampling (update) region, and obtained a closed-form expression for the value function of the bandit's Bellman equation. 
	\item [3)] We proved, with the monotonicity condition and the closed-form value function, that monotonic bandits ($p+q<1$) and oscillating bandits ($p+q>1$) are both indexable, hence establishing the RMAB's indexability and the viability of a Whittle index policy.
	\item [4)] We further developed an algorithm to compute the Whittle index.
\end{itemize}
We also studied a special case where $p=q$, and showed that the closed-form expression of Whittle index is available in this case. We remark that the results in this paper are valid not just for the UoI problem of focus here. Specifically, our results and methods are valid for a class of RMABs wherein the bandits are binary Markov processes, and the penalties are concave functions (not limited to UoI) of the bandits' belief states.
We demonstrated the excellent performance of our Whittle index policy for these RMABs by numerical results (UoI problems as well as other problems that fit into the class).

\section*{Appendix A}  \label{Apd:A}
\subsection{Proof of Lemma \hyperref[Lem:2]{2}}
\begin{IEEEproof}
	Let $\kappa_1$ denote the active set of an arbitrary  policy $\pi $, i.e., $\kappa_1$ is the set of belief states in which the active action is taken under policy $\pi$. If $\kappa_1$ is empty, all belief states will evolve to $\omega^*$; hence the policy is unichain. We then consider the case that $\kappa_1$  is not empty.
	
	First, if $\omega^* \in \kappa_1$, then any $\omega\in [0,1]$ can evolve to a belief state which also belongs to the active set $\kappa_1$  within finite time, hence it can evolve to $p$ and $1-q$ at some point (note: recall our assumption that the belief states in the neighborhood of $\omega^*$ also belongs to $\kappa_1$ if $\omega^* \in \kappa_1$). In this case, the policy is unichain.
	
	Second, assume $\omega^* \notin \kappa_1$ and $p^{(n)}\notin \kappa_1$ for all $n$. Any  $\omega \in \kappa_1$ would transit in the next step to belief state $p$ with probability $1-\omega$, and since $p^{(n)}\notin \kappa_1$ for all $n$, the belief state will eventually evolve to $\omega^*$.  On the other hand, any $\omega \notin \kappa_1$ will either evolve toward $\omega^*$ without an active action being taken in future steps, or evolve to a belief state in $\kappa_1$ (in which an active action is taken), the result of the previous sentence means that eventually the belief state will evolve toward $\omega^*$. 
	Hence the policy is unichain. Likewise, if $\omega^* \notin \kappa_1$ and $1-q^{(n)}\notin \kappa_1$ for all $n$, the policy is also a unichain.
	
	Finally, if $\omega^* \notin \kappa_1$ and $p^{(n)}\in \kappa_1, 1-q^{(m)}\in \kappa_1$ for some $n$ and $m$ respectively, then policy $\pi$ partitions the belief state space into two chains: 1) chain 1 consists of belief states that can evolve to $p$ and $1-q$; 2) chain 2 consists of belief states that evolve to $\omega^*$. Hence the policy is multichain.
	
	In summary, only the last case leads to a multichain policy, and the multichain policy is a two-chain policy. We thus can obtain the properties as summarized in Lemma \hyperref[Lem:2]{2}.
\end{IEEEproof}

\subsection{Proof of Lemma \hyperref[Lem:3]{3}}
\begin{IEEEproof}
	Suppose that $\pi$ is a multichain policy with active set $\kappa_1$. Then $\omega^*\notin \kappa_1$. As stated in Lemma \hyperref[Lem:2]{2}, let $\Omega_1$ denote chain 1 that includes $p$ and $1-q$, and $\Omega_2$ denote chain 2 that includes $\omega^*$. Let ${g^\pi }({\Omega _i})$ denote the average penalty obtained by this policy conditioned on an initial state belonging to $\Omega_i, i=1,2$. According to the Bellman equation, for  $\Omega_2$ that includes $\omega^*$, we have
	\begin{align*}
	V({\omega ^ * }) + {g^\pi }({\Omega _2}) = H({\omega ^ * }) + V\left( {\tau ({\omega ^ * })} \right) = H({\omega ^ * }) + V\left( {{\omega ^ * }} \right),
	\end{align*}
	which means that ${g^\pi }({\Omega _2}) = H({\omega ^ * })$. Likewise, for $\Omega_1$ that includes  $p$ and  $1-q$, we have
	\begin{align} \label{eq:lemma3-1}
	&V(p)  = \sum\limits_{i = 1}^L {\left[ {H\left( {{p^{(i)}}} \right) - {g^\pi }({\Omega _1})} \right]} + \lambda  + {p^{(L)}}V(1 - q)  \\ \label{eq:lemma3-2}
	&{q^{(K)}}V(1 - q) =  \sum\limits_{i = 1}^K {\left[ {H\left( {{q^{(i)}}} \right) - {g^\pi }({\Omega _1})} \right]}  + \lambda,
	\end{align}
	where integer $L \triangleq \min \{ n:{\tau ^n}(p) \in {\kappa _1}\} $ and integer $K \triangleq \min \{ n:{\tau ^n}(1 - q) \in {\kappa _1}\} $. It follows from \eqref{eq:lemma3-1} and \eqref{eq:lemma3-2} that
	 \begin{align} \label{eq:lemma3-g}
	 {g^\pi }({\Omega _1}) = \frac{{{q^{(K)}}\left[ {\sum\limits_{i = 1}^L {H\left( {{p^{(i)}}} \right)}  + \lambda } \right] + {p^{(L)}}\left[ {\sum\limits_{i = 1}^K {H\left( {{q^{(i)}}} \right)}  + \lambda } \right]}}{{L{q^{(K)}} + K{p^{(L)}}}}.
	 \end{align}
	 Given $\lambda$, ${g^\pi }({\Omega _1})$ only depends on $L$ and $K$. If ${g^\pi }({\Omega _2}) \le {g^\pi }({\Omega _1})$, consider the never-sample policy that takes passive action in all belief states, denoted by $o$. We have
	 \begin{align*}
	 {g^o}({\Omega _2}) = {g^o}({\Omega _1}) = H({\omega ^ * }) = {g^\pi }({\Omega _2}) \le {g^\pi }({\Omega _1}),
	 \end{align*}
	 where ${g^o}({\Omega _i})$ is the average penalty obtained by policy $o$  with an initial state that  belongs to $\Omega_i,i=1,2$. In this case, policy $o$ is not worse than $\pi$ for any initial state. Note that $o$ is a unichain policy.
	 
	 On the other hand, if ${g^\pi }({\Omega _2}) > {g^\pi }({\Omega _1})$, we need to consider two cases. First, if $p+q<1$, then ${p^{(L)}} < {\omega ^ * } < 1 - {q^{(K)}}$. Consider a policy $\varsigma$ that takes active action in  $\omega\in [{p^{(L)}},1 - {q^{(K)}}]$. Policy $\varsigma$ is unichain, and its average penalty is given by \eqref{eq:lemma3-g}. Then we have
	 \begin{align*}
	 {g^\varsigma }({\Omega _2}) = {g^\varsigma }({\Omega _1}) = {g^\pi }({\Omega _1}) < {g^\pi }({\Omega _2}),
	 \end{align*}
	 which means that policy $\varsigma$ is better than policy $\pi$.
	 
	 Second, if $p+q>1$, the order of ${p^{(L)}},1 - {q^{(K)}}$ and $\omega^*$ has 6 possibilities. For each case, we can construct a policy $\varsigma'$ with active set ${\kappa _1} \cup {\cal S}$, where
	 \begin{align*}
	 \cal S= \begin{cases}
	 [x,\omega^*+\epsilon], & \text{if } x<\omega^* \\
	 [\omega^* - \epsilon,x], & \text{if } x > \omega^*
	 \end{cases}
	 \end{align*}
	 with $x=\min \{p^{(L)}, 1-q^{(K)}\}$, and $\epsilon>0$ is small enough such that policy  $\varsigma'$ takes passive action in belief states ${p^{(n)}},n < L$ and $1 - {q^{(m)}},m < K$. Then, policy $\varsigma'$ is a unichain policy. We have
	 \begin{align*}
	 {g^{\varsigma '}}({\Omega _2}) = {g^{\varsigma '}}({\Omega _1}) = {g^\pi }({\Omega _1}) < {g^\pi }({\Omega _2}),
	 \end{align*}
	 which means that policy $\varsigma '$ is better than policy $\pi$.
	 
	 In summary, for an arbitrary multichain policy, we can find a unichain policy that is not worse than it. Therefore, we conclude that the single-bandit problem can be optimized by a unichain policy.
\end{IEEEproof}

\subsection{Proof of Lemma \hyperref[Lem:5]{5}}
\begin{IEEEproof}
	Statement 1 follows immediately from the Bellman equation. We prove statement 2 by contradiction. First, consider the case of $p+q<1$. Assume there exists a $\lambda$ such that $\omega^l < \omega^u < \omega^*$, then for any $\omega\in (\omega^l,\omega^u)$, we have
	\begin{align} \notag
	a(\omega ) &= \omega V(1 - q) + \lambda  < r(\omega ) = V(\tau (\omega )) \\
	&\le H(\tau (\omega )) - g + \tau (\omega )V(1 - q) + \lambda. 
	\end{align}
	Since $\omega < \omega^*$ and $p+q<1$, $\omega<\tau(\omega)=p+\omega (1-p-q)$, then the above inequality implies that
	\begin{align*}
	V(1 - q) > \frac{{H(\tau (\omega )) - g}}{{\omega  - \tau (\omega )}}.
	\end{align*}
	For $\omega\in (0,\omega^*)$ and $\tau = \tau(\omega)$, define a function $f(\omega ) \triangleq [H(\tau)-g]/(\omega -\tau)$, then we can compute its derivative as follows:
	\begin{align*}
	f'(\omega ) = \frac{{H'(\tau )(1 - p - q)(\omega  - \tau ) - \left[ {H(\tau ) - g} \right](p + q)}}{{{{\left( {\omega  - \tau } \right)}^2}}}.
	\end{align*}
	Further, define 
	\begin{align*}
	h(\omega ) \triangleq H'(\tau )(1 - p - q)(\omega  - \tau ) - \left[ {H(\tau ) - g} \right](p + q).
	\end{align*}
	We have
	\begin{align} \label{eq:lemma5-2}
	h'(\omega ) = H''(\tau ){(1 - p - q)^2}(\omega  - \tau ) \ge 0.
	\end{align}
	The inequality follows from the concavity of $H(\tau)$. Note that the average penalty obtained by the never-sample policy is $H(\omega^*)$, hence the optimal average penalty $g\le H(\omega^*)$. Then \eqref{eq:lemma5-2} means that $h(\omega ) \le h({\omega ^ * }) =  - \left[ {H({\omega ^ * }) - g} \right](p + q) \le 0$, hence $f'(\omega)\le0$. We thus have
	\begin{align*}
	V(1 - q) > \frac{{H(\tau (\omega )) - g}}{{\omega  - \tau (\omega )}} = f(\omega ) \ge f(\tau  ) = \frac{{H({\tau ^2}(\omega )) - g}}{{\tau (\omega ) - {\tau ^2}(\omega )}},
	\end{align*}
	which implies that $a(\tau)<r(\tau)$. Therefore, for any $\omega  \in ({\omega ^l},{\omega ^u})$, the optimal action in belief state $\tau(\omega)$ should also be the active action, i.e., $\tau (\omega ) \in ({\omega ^l},{\omega ^u})$. If $\omega^u < \omega^*$, then there must exist an $\omega  \in ({\omega ^l},{\omega ^u})$ such that $\tau (\omega ) \notin ({\omega ^l},{\omega ^u})$, which leads to a contradiction. 
	
	On the other hand, if $p+q>1$. Assume there exists a $\lambda$ such that $\omega^l < \omega^u < \omega^*$, then for any $\omega\in (\omega^l,\omega^u)$, we have  $\tau\notin (\omega^l,\omega^u)$. Hence 
	\begin{align*}
	&\omega V(1 - q) + \lambda  < V(\tau ) = H(\tau ) - g + V({\tau ^2}(\omega ))\\
	\le& H(\tau ) + H({\tau ^2}(\omega )) - 2g + {\tau ^2}(\omega )V(1 - q) + \lambda. 
	\end{align*}
	Then we have
	\begin{align*}
	V(1 - q) > \frac{{H(\tau ) + H({\tau ^2}(\omega )) - 2g}}{{\omega  - {\tau ^2}(\omega )}} \triangleq {f_0}(\omega ).
	\end{align*}
	Applying a similar method as in the case of $p+q<1$, we can verify that $f'_0(\omega)\le0$. Hence
	\begin{align*}
	V(1 - q) &> \frac{{H(\tau ) + H({\tau ^2}(\omega )) - 2g}}{{\omega  - {\tau ^2}(\omega )}} \\
	&\ge \frac{{H({\tau ^2}(\omega )) + H({\tau ^4}(\omega )) - 2g}}{{{\tau ^2}(\omega ) - {\tau ^4}(\omega )}}.
	\end{align*}
	From the above inequality, we have
	\begin{align*}
	a\left( {{\tau ^2}(\omega )} \right) < r\left( {{\tau ^2}(\omega )} \right).
	\end{align*}
	Therefore, for any $\omega  \in ({\omega ^l},{\omega ^u})$, we have $\tau^2 (\omega ) \in ({\omega ^l},{\omega ^u})$. Hence $\omega^u$ can not be smaller than $\omega^*$. This completes the proof.
\end{IEEEproof}

\section*{Appendix B}  \label{apd:C}
\subsection*{A. Proof of Lemma \hyperref[Lem:8]{8}}
\begin{IEEEproof}
	Since $p+q>1$, as stated in Lemma 1, $\tau^k(\tau)$ is an oscillating function of $k$. We have
	\begin{align} \label{eq:lemma8-1}
	\frac{{{\tau ^k}(\omega ) - {\omega ^ * }}}{{\omega  - {\omega ^ * }}}  = \begin{cases}
	{(p + q - 1)^{2n}}, &\text{if }k = 2n\\
	- {(p + q - 1)^{2n + 1}}, & \text{if }k = 2n + 1
	\end{cases}
	\end{align} 
	where $n\in \mathbb{N}$. If ${\omega ^ * } \in ({\omega ^l},{\omega ^u})$, both $\tau^{2n}(\omega)$ and $\tau^{2n+1}(\omega)$ can enter the sampling region; while if ${\omega ^ * } \notin ({\omega ^l},{\omega ^u})$, either $\tau^{2n}(\omega)$ or $\tau^{2n+1}(\omega)$ is possible to enter the sampling region. In both cases, if $\omega  \in ({\omega ^l},{\omega ^u})$ then $T(\omega ,{\omega ^l},{\omega ^u}) = 0$ by definition. We then consider the cases that $\omega  \notin ({\omega ^l},{\omega ^u})$.
	
	First, if ${\omega ^ * } \in ({\omega ^l},{\omega ^u})$ and $\omega\ge \omega^u$, then $\tau^{2n}(\omega)$ is decreasing w.r.t. $n$ and $\tau^{2n}(\omega)>\omega^*$; while  $\tau^{2n+1}(\omega)$ is increasing w.r.t. $n$ and $\tau^{2n+1}(\omega)<\omega^*$. Applying a similar method as in the proof of Lemma \hyperref[Lem:7]{7}, we can obtain that the minimum integer  $n$ such that ${\tau ^{2n}}(\omega ) \in ({\omega ^l},{\omega ^u})$ is given by
	\begin{align} \label{eq:lemma8-2}
	{n_1} = \left\lfloor {\frac{1}{2}{{\log }_{p + q - 1}}\frac{{{\omega ^u} - {\omega ^ * }}}{{\omega  - {\omega ^ * }}}} \right\rfloor  + 1 = \left\lfloor {\varphi ({\omega ^u},\omega )} \right\rfloor  + 1.
	\end{align}
	On the other hand, the minimum integer $n$ such that ${\tau ^{2n+1}}(\omega ) \in ({\omega ^l},{\omega ^u})$ is given by
	\begin{align} \label{eq:lemma8-3}
	{n'_1} = \left\lfloor {\frac{1}{2}\left( {{{\log }_{p + q - 1}}\frac{{{\omega ^l} - {\omega ^ * }}}{{{\omega ^ * } - \omega }} - 1} \right)} \right\rfloor  + 1 = \left\lfloor {\phi ({\omega ^l},\omega )} \right\rfloor  + 1.
	\end{align}
	Then the hitting time of $\omega$ is the smaller integer between $2n_1$ and $2n'_1 +1$, i.e., 
	\begin{align} \label{eq:lemma8-4}
	T(\omega ,{\omega ^l},{\omega ^u}) = \min \left\{ {2\left\lfloor {\phi ({\omega ^l},\omega )} \right\rfloor  + 3,2\left\lfloor {\varphi ({\omega ^u},\omega )} \right\rfloor  + 2} \right\}.
	\end{align}
	
	Second, if ${\omega ^ * } \in ({\omega ^l},{\omega ^u})$ and $\omega\le \omega^l$, then $\tau^{2n}(\omega)$ is increasing w.r.t. $n$ and $\tau^{2n}(\omega)<\omega^*$; while  $\tau^{2n+1}(\omega)$ is decreasing w.r.t. $n$ and $\tau^{2n+1}(\omega)>\omega^*$. Applying a similar argument as above, we can obtain the hitting time as follows:
	\begin{align} \label{eq:lemma8-5}
	T(\omega ,{\omega ^l},{\omega ^u}) = \min \left\{ {2\left\lfloor {\phi ({\omega ^u},\omega )} \right\rfloor  + 3,2\left\lfloor {\varphi ({\omega ^l},\omega )} \right\rfloor  + 2} \right\}.
	\end{align}
	Putting \eqref{eq:lemma8-4} and \eqref{eq:lemma8-5} together gives the expression of $T(\omega ,{\omega ^l},{\omega ^u})$ in the case of  ${\omega ^ * } \in ({\omega ^l},{\omega ^u})$.
	
	If $\omega^u=\omega^*$, \eqref{eq:lemma8-4} and \eqref{eq:lemma8-5} are still valid. In this case, ${\varphi ({\omega ^u},\omega )}= {\phi ({\omega ^u},\omega )}=\infty$, and the hitting time depends on $\omega^l$.  The case that $\omega^l=\omega^*$ is similar. We then consider the case of ${\omega ^ * } \notin [{\omega ^l},{\omega ^u}]$. Since $\omega^u \ge \omega^*$ for any $\lambda$, the only possibility in this case is $\omega^l>\omega^*$. The belief state space is divided into four parts: (i) $\{ \omega :\omega  \ge {\omega ^u}\} $; (ii) $({\omega ^l},{\omega ^u})$; (iii) $\{ \omega :\omega  \le {\omega ^l},\tau (\omega ) \le {\omega ^l}\} $; (iv) $\{ \omega :\omega  \le {\omega ^l},\tau (\omega ) > {\omega ^l}\} $. It is easy to verify that $T(\omega ,{\omega ^l},{\omega ^u}) = 0$ for $\omega$ in the second part and $T(\omega ,{\omega ^l},{\omega ^u}) = \infty$ for $\omega$ in the third part. We next focus on the remaining two parts.
	
	For any $\omega\ge \omega^u$, ${\tau ^{2n + 1}}(\omega ) < {\omega ^ * } \le {\omega ^l}$ for all $n$. Hence only the even branch, $\tau^{2n}(\omega)$, is possible to enter the sampling region. The hitting time of $\omega$  is an even integer $k=2n$ such that ${\tau ^{k - 2}}(\omega ) \ge {\omega ^u} > {\tau ^k}(\omega ) > {\omega ^l}$.  As discussed above, the integer that satisfies ${\tau ^{2n - 2}}(\omega ) \ge {\omega ^u} > {\tau ^{2n}}(\omega )$ is $n_1$ given by \eqref{eq:lemma8-2}. Likewise, the integer that satisfies ${\tau ^{2n}}(\omega ) > {\omega ^l} \ge {\tau ^{2n + 2}}(\omega )$ is given by
	\begin{align}
	{m_1} = \left\lceil {\frac{1}{2}{{\log }_{p + q - 1}}\frac{{{\omega ^l} - {\omega ^ * }}}{{\omega  - {\omega ^ * }}}} \right\rceil  - 1 = \left\lceil {\varphi ({\omega ^l},\omega )} \right\rceil  - 1.
	\end{align}
	Note that $2m_1$ is the maximum integer such that ${\tau ^{2{m_1}}}(\omega ) > {\omega ^l}$. Then, $T(\omega ,{\omega ^l},{\omega ^u}) = 2{n_1}$ if $n_1 \le m_1$. While if $n_1>m_1$, we can not find an integer to make ${\tau ^k}(\omega ) \in ({\omega ^l},{\omega ^u})$; hence $T(\omega ,{\omega ^l},{\omega ^u}) = \infty$.
	
	For $\omega$ in part (iv), ${\tau ^{2n}}(\omega ) < {\omega ^ * } \le {\omega ^l}$ for all $n$. Hence only $\tau^{2n+1}(\omega)$ is possible to enter the sampling region. In this case, the hitting time of $\omega$ is an odd integer $k=2n+1$ such that ${\tau ^{k - 2}}(\omega ) \ge {\omega ^u} > {\tau ^k}(\omega ) > {\omega ^l}$. Similar to the previous case, we first find the integer that satisfies ${\tau ^{2n - 1}}(\omega ) \ge {\omega ^u} > {\tau ^{2n + 1}}(\omega )$, which is given by
	\begin{align*}
	{n_2} = \left\lfloor {\frac{1}{2}\left( {{{\log }_{p + q - 1}}\frac{{{\omega ^u} - {\omega ^ * }}}{{{\omega ^ * } - \omega }} - 1} \right)} \right\rfloor  + 1 = \left\lfloor {\phi ({\omega ^u},\omega )} \right\rfloor  + 1.
	\end{align*}
	Likewise, the integer that satisfies ${\tau ^{2n + 1}}(\omega ) > {\omega ^l} \ge {\tau ^{2n + 3}}(\omega )$ is given by
	\begin{align*}
	{m_2} = \left\lceil {\frac{1}{2}\left( {{{\log }_{p + q - 1}}\frac{{{\omega ^l} - {\omega ^ * }}}{{{\omega ^ * } - \omega }} - 1} \right)} \right\rceil  - 1 = \left\lceil {\phi ({\omega ^l},\omega )} \right\rceil  - 1.
	\end{align*}
	Then, $T(\omega ,{\omega ^l},{\omega ^u}) = 2{n_2} + 1$ if $n_2\le m_2$. If $n_2>m_2$, we can not find an integer $k$ such that ${\tau ^k}(\omega ) \in ({\omega ^l},{\omega ^u})$; hence $T(\omega ,{\omega ^l},{\omega ^u}) = \infty$.
	
	Putting the above results together gives the expression of $T(\omega ,{\omega ^l},{\omega ^u})$ presented in Lemma \hyperref[Lem:8]{8}.
\end{IEEEproof}

\section*{Appendix C} \label{apd:D}
\textit{\textbf{Lemma} C1:} Let $L = T(p,{\omega ^l},{\omega ^u}) + 1$ and $K = T(1 - q,{\omega ^l},{\omega ^u}) + 1$, where $T(\bullet,{\omega ^l},{\omega ^u})$ is given by Lemma \hyperref[Lem:7]{7}. Define a function
\begin{align*}
{f_\delta }(\omega ) \triangleq (L - K)\left[ {p - \omega (p + q)} \right] - {q^{(K)}} - {p^{(L)}}.
\end{align*}
If $p+q<1$ and ${\omega ^ * } = p/(p + q) \in ({\omega ^l},{\omega ^u})$, then ${f_\delta }({\omega ^l}) < 0,{f_\delta }({\omega ^u}) < 0$. 
\begin{IEEEproof}
	Since $\omega^l < \omega^*$, then $p - {\omega ^l}(p + q) > 0$. According to the definition of hitting time, we have $K \ge 1,{p^{(L)}} > {\omega ^l}$ and $\left( {1 - {q^{(K)}}} \right) < {\omega ^u}$. Then
	\begin{align*}
	{f_\delta }({\omega ^l}) < (L - 1)\left( {p - {\omega ^l}(p + q)} \right) - \left( {1 - {\omega ^u}} \right) - {\omega ^l}.
	\end{align*}
	If $T(p,{\omega ^l},{\omega ^u}) = 0$, then $L=1$ and ${f_\delta }({\omega ^l}) <  - \left( {1 - {\omega ^u}} \right) - {\omega ^l} < 0$. If $T(p,{\omega ^l},{\omega ^u}) > 0$, according to Lemma 7, we have
	\begin{align*}
	&{f_\delta }({\omega ^l}) < (L - 1)\left( {p - {\omega ^l}(p + q)} \right) - \left( {1 - {\omega ^u}} \right) - {\omega ^l}\\
	\le& \left( {{{\log }_{1 - p - q}}\frac{{{\omega ^l} - {\omega ^ * }}}{{p - {\omega ^ * }}} + 1} \right)\left[ {p - {\omega ^l}(p + q)} \right] - {\omega ^l} - \left( {1 - {\omega ^u}} \right).
	\end{align*}
	For $x\in(0,\omega^*)$, define the following function,
	\begin{align*}
	h_1(x) = \left( {{{\log }_{1 - p - q}}\frac{{x - {\omega ^ * }}}{{p - {\omega ^ * }}} + 1} \right)\left[ {p - x(p + q)} \right] - x.
	\end{align*}
	We can obtain the derivative of $h(x)$ as follows
	\begin{align*} \notag
	h_1'(x) = & - \left( {{{\log }_{1 - p - q}}\frac{{x - {\omega ^ * }}}{{p - {\omega ^ * }}} + 1} \right)(p + q) \\
	& - \frac{{p + q{\rm{ + }}\ln (1 - p - q)}}{{\ln (1 - p - q)}}.
	\end{align*}
	Since $p\le q$, we can verify that the first term of $h_1'(x)$ (include the ``$-$'') is non-positive for all $x\in(0,\omega^*)$. For the second term, let 
	\begin{align*}
	l(z) = z - 1 - \ln z,
	\end{align*}
	where $z = 1 - p - q \in (0,1)$. It is easy to verify that $l'(z)<0$; hence $l(z)>l(1)=0$. We then have
	\begin{align} \label{eq:LD2-1}
	 - \frac{{p + q{\rm{ + }}\ln (1 - p - q)}}{{\ln (1 - p - q)}} = \frac{{l(z)}}{{\ln z}} < 0.
	\end{align}
	Hence $h_1'(x)<0$ and $h_1(x)<h_1(0)=0$. Therefore, ${f_\delta }({\omega ^l}) < h_1({\omega ^l}) - \left( {1 - {\omega ^u}} \right) < 0$.
	
	On the other hand, $\omega^u>\omega^*$ means that $p - {\omega ^u}(p + q) < 0$. Since $L\ge 1$, we have
	\begin{align*}
	{f_\delta }({\omega ^u}) < (1 - K)\left( {p - {\omega ^u}(p + q)} \right) - \left( {1 - {\omega ^u}} \right) - {\omega ^l}.
	\end{align*}
	If $T(1 - q,{\omega ^l},{\omega ^u}) = 0$, then $K=1$ and
	\begin{align*}
	{f_\delta }({\omega ^u}) <  - \left( {1 - {\omega ^u}} \right) - {\omega ^l} < 0.
	\end{align*}
	 If $T(1 - q,{\omega ^l},{\omega ^u}) > 0$, we have
	 \begin{align*}
	 {f_\delta }({\omega ^u}) <& (1 - K)\left( {p - \omega ^u (p + q)} \right) - \left( {1 - {\omega ^u}} \right) - {\omega ^l}\\
	 <&  - \left( {{{\log }_{1 - p - q}}\frac{{{\omega ^u} - {\omega ^ * }}}{{1 - q - {\omega ^ * }}} + 1} \right)\left[ {p - {\omega ^u}(p + q)} \right] \\
	 & - \left( {1 - {\omega ^u}} \right) - {\omega ^l}.
	 \end{align*}
	 For $x\in (\omega^*,1)$, define a function
	  \begin{align*}
	  {h_2}(x) =  \left( {{{\log }_{1 - p - q}}\frac{{x - {\omega ^ * }}}{{1 - q - {\omega ^ * }}} + 1} \right)\left[ { x(p + q)-p} \right] - \left( {1 - x} \right).
	  \end{align*}
	  Applying a similar method as in the case of $h_1(x)$, we can verify that $h'_2(x)>0$ for all $x\in (\omega^*,1)$. Hence $h_2(x)<h_2(1)=0$. We thus conclude that ${f_\delta }({\omega ^u}) < {h_2}({\omega ^u}) - {\omega ^l} < 0$.
	
\end{IEEEproof}

\textit{\textbf{Lemma} C2:} Let $L = T(p,{\omega ^l},{\omega ^u}) + 1$ and $K = T(1 - q,{\omega ^l},{\omega ^u}) + 1$, where $T(\bullet,{\omega ^l},{\omega ^u})$ is given by Lemma \hyperref[Lem:8]{8}. Define a function
\begin{align*}
{f_1 }(\omega ) \triangleq (L - K)\left[ {p - \omega (p + q)} \right] - {q^{(K)}} - {p^{(L)}}.
\end{align*}
If $p+q>1$ and ${\omega ^ * } = p/(p + q) \in ({\omega ^l},{\omega ^u})$, then ${f_1 }({\omega ^l}) < 0,{f_1 }({\omega ^u}) < 0$. 
\begin{IEEEproof}
	Since $\omega^l < \omega^*$, then $p - {\omega ^l}(p + q) > 0$. Note that $K \ge 1,{p^{(L)}} > {\omega ^l}$, and $1 - {q^{(K)}} < {\omega ^u}$. Then
	\begin{align*}
	{f_1}({\omega ^l}) < (L - 1)\left( {p - {\omega ^l}(p + q)} \right) - \left( {1 - {\omega ^u}} \right) - {\omega ^l}.
	\end{align*}
	If $T(p,{\omega ^l},{\omega ^u}) = 0$, then $L=1$ and
	\begin{align*}
	{f_1 }({\omega ^l}) <  - \left( {1 - {\omega ^u}} \right) - {\omega ^l} < 0.
	\end{align*} 
	If $T(p,{\omega ^l},{\omega ^u}) > 0$, then $p\ge\omega^u$ (note that $p+q>1$ implies $p>\omega^*$).  According to $T(\bullet,{\omega ^l},{\omega ^u})$ in Lemma 8, we have
	\begin{align*}
	L &\le {2\left\lfloor {\frac{1}{2}\left( {{{\log }_{p + q - 1}}\frac{{{\omega ^l} - {\omega ^ * }}}{{{\omega ^ * } - p}} - 1} \right)} \right\rfloor  + 3}  + 1\\
	&\le {\log _{p + q - 1}}\frac{{{\omega ^l} - {\omega ^ * }}}{{{\omega ^ * } - p}} + 3.
	\end{align*}
	Therefore, 
	\begin{align*}
	{f_1}({\omega ^l}) <& \left( {{{\log }_{p + q - 1}}\frac{{{\omega ^l} - {\omega ^ * }}}{{{\omega ^ * } - p}} + 2} \right)\left( {p - {\omega ^l}(p + q)} \right)\\
	& - \left( {1 - {\omega ^u}} \right) - {\omega ^l}.
	\end{align*}
	For any $x\in [0,\omega^*)$, define a function
	\begin{align*}
	h_3(x) = \left( {{{\log }_{p + q - 1}}\frac{{x - {\omega^*}}}{{{\omega^*} - p}} + 2} \right)\left( {p - x(p + q)} \right) - x.
	\end{align*}
	The derivative of $h(x)$ is 
	\begin{align*}
	h_3'(x) =  &- \left( {{{\log }_{p + q - 1}}\frac{{x - {\omega ^ * }}}{{{\omega ^ * } - p}} + 2} \right)(p + q) \\
	&- \frac{{p + q{\rm{ + }}\ln (p + q - 1)}}{{\ln (p + q - 1)}}.
	\end{align*}
	The first term of $h_3'(x)$ (include the ``$-$'') is negative for all  $x\in [0,\omega^*)$. Since $p+q-1 \in (0,1)$, applying a similar method as in the proof of Lemma C1, just around \eqref{eq:LD2-1}, we can prove that the second term of  $h_3'(x)$ is also negative. Hence ${f_1}({\omega ^l}) < h({\omega ^l}) - (1 - {\omega ^u}) < 0$.
	
	We then show that ${f_1}({\omega ^u}) < 0$. Note that $p - {\omega ^u}(p + q) < 0$ because $\omega^u > \omega^*$. Since $L\ge 1$, we have
	\begin{align*}
	{f_1}({\omega ^u}) < (1 - K)\left( {p - {\omega ^u}(p + q)} \right) - \left( {1 - {\omega ^u}} \right) - {\omega ^l}.
	\end{align*}
	If $T(1 - q,{\omega ^l},{\omega ^u}) = 0$, then $K=1$ and
	\begin{align*}
	{f_1 }({\omega ^u}) <  - \left( {1 - {\omega ^u}} \right) - {\omega ^l} < 0.
	\end{align*}
	If $T(1 - q,{\omega ^l},{\omega ^u}) > 0$, then we must have $1-q\le \omega^l$. Then according to Lemma 8,
	\begin{align*}
	K &\le {2\left\lfloor {\frac{1}{2}\left( {{{\log }_{p + q - 1}}\frac{{{\omega ^u} - {\omega ^ * }}}{{{\omega ^ * } - (1 - q)}} - 1} \right)} \right\rfloor  + 3}  + 1\\
	&\le {\log _{p + q - 1}}\frac{{{\omega ^u} - {\omega ^ * }}}{{{\omega ^ * } - (1 - q)}} + 3.
	\end{align*}
	Therefore,
	\begin{align*}
	{f_1}({\omega ^u}) <  &- \left( {{{\log }_{p + q - 1}}\frac{{{\omega ^u} - {\omega ^ * }}}{{{\omega ^ * } - (1 - q)}} + 2} \right)\left( {p - {\omega ^u}(p + q)} \right) \\
	& - \left( {1 - {\omega ^u}} \right) - {\omega ^l}.
	\end{align*}
	For $x\in (\omega^*,1]$, define
	\begin{align*}
	{h_4}(x) =& - \left( {{{\log }_{p + q - 1}}\frac{{x - {\omega ^ * }}}{{{\omega ^ * } - (1 - q)}} + 2} \right)\left( p-{ x(p + q)} \right) \\
	& - \left( {1 - x} \right).
	\end{align*}
	Applying a similar method as in the case of $h_3(x)$, we can verify that $h_4'(x)>0$ and $h_4(x)<h_4(1)=0$. We thus conclude that ${f_1}({\omega ^u}) < h_4({\omega ^u}) - {\omega ^l} < 0$.
\end{IEEEproof}

\textit{\textbf{Lemma} C3:} Let $(\omega^l,\omega^u)$ denote the sampling region of the optimal policy for an oscillating bandit ($p+q>1$) with a service charge $\lambda$. For any $\lambda>0$ such that $T(p,{\omega ^l},{\omega ^u}) = \infty $ or $T(1-q,{\omega ^l},{\omega ^u}) = \infty $, and $\omega^l\ge \omega^*$, the monotonicity condition holds:
\begin{align*} 
\frac{{\partial a({\omega ^l},\lambda )}}{{\partial \lambda }} > \frac{{\partial r({\omega ^l},\lambda )}}{{\partial \lambda }}\text{ and }\frac{{\partial a({\omega ^u},\lambda )}}{{\partial \lambda }} > \frac{{\partial r({\omega ^u},\lambda )}}{{\partial \lambda }}.
\end{align*}
\begin{IEEEproof}
	Let $L = T(p,{\omega ^l},{\omega ^u}) + 1$ and $K = T(1 - q,{\omega ^l},{\omega ^u}) + 1$. We have three cases: (1) $L=\infty, K<\infty$; (2) $L<\infty, K=\infty$; (3) $L=\infty,K=\infty$.
	
	First, if $L=\infty$ and $K<\infty$. According to the Bellman equation
	\begin{align} \label{eq:LD3-1}
	&V(p) = \mathop {\lim }\limits_{L \to \infty } \left\{ {\sum\limits_{k = 1}^L {\left[ {H\left( {{p^{(k)}}} \right) - g} \right]}  + V\left( {{p^{(L)}}} \right)} \right\} = 0 \\ \notag
	&V(1 - q) = \sum\limits_{k = 1}^K {\left[ {H\left( {{q^{(k)}}} \right) - g} \right]}  + \lambda  + \left( {1 - {q^{(K)}}} \right)V(1 - q).
	\end{align}
	Since $\omega^l\ge \omega^*$, we have $T({\omega ^l},{\omega ^l},{\omega ^u}) = \infty $, then
	\begin{align} \label{eq:LD3-2}
	r({\omega ^l},\lambda ) =& \mathop {\lim }\limits_{k \to \infty } \sum\limits_{i = 1}^k {\left[ {H\left( {{\tau ^i}({\omega ^l})} \right) - g} \right]} \nonumber \\
	 &- \mathop {\lim }\limits_{k \to \infty } \sum\limits_{i = 1}^k {\left[ {H\left( {{p^{(i)}}} \right) - g} \right]}. 
	\end{align}
	Note that $g=H(\omega^*)$, then $\partial g/\partial \lambda=0$. We thus have
	\begin{align*}
	\frac{{\partial a({\omega ^l},\lambda )}}{{\partial \lambda }}  = \frac{{{\omega ^l}}}{{{q^{(K)}}}} + 1 > \frac{{\partial r({\omega ^l},\lambda )}}{{\partial \lambda }} = 0.
	\end{align*}
	
	For $\omega^u$, if ${\tau ^2}({\omega ^u}) \in ({\omega ^l},{\omega ^u})$, then $T({\omega ^u},{\omega ^l},{\omega ^u}) = 2$, otherwise, $T({\omega ^u},{\omega ^l},{\omega ^u}) = \infty$. For the former case, $r(\omega^u,\lambda)$ is given by
	\begin{align*}
	r({\omega ^u},\lambda ) = \sum\limits_{i = 1}^2 {\left[ {H\left( {{\tau ^i}({\omega ^u})} \right) - g} \right]}  + a\left( {{\tau ^2}({\omega ^u}),\lambda } \right).
	\end{align*}
	Hence the partial derivative of $r(\omega^u,\lambda)$ is
	\begin{align*}
	\frac{{\partial r({\omega ^u},\lambda )}}{{\partial \lambda }}  = \frac{{{\tau ^2}({\omega ^u})}}{{{q^{(K)}}}} + 1 < \frac{{\partial a({\omega ^u},\lambda )}}{{\partial \lambda }} = \frac{{{\omega ^u}}}{{{q^{(K)}}}} + 1.
	\end{align*}
	If $T({\omega ^u},{\omega ^l},{\omega ^u}) = \infty$, then $r(\omega^u,\lambda)$ is with the same form as $r(\omega^l,\lambda)$ given by \eqref{eq:LD3-2}. Hence
	\begin{align*}
	\frac{{\partial r({\omega ^u},\lambda )}}{{\partial \lambda }} = 0 < \frac{{\partial a({\omega ^u},\lambda )}}{{\partial \lambda }} = \frac{{{\omega ^u}}}{{{q^{(K)}}}} + 1.
	\end{align*}
	
	Second, if $L<\infty$ and $K=\infty$. According to the Bellman equation,
	\begin{align} \notag
	&V(p) = \sum\limits_{k = 1}^L {\left[ {H\left( {{p^{(k)}}} \right) - g} \right]}  + \lambda  + {p^{(L)}}V(1 - q) = 0, \\ \label{eq:LD3-3}
	&V(1 - q) = \mathop {\lim }\limits_{K \to \infty } \left\{ {\sum\limits_{k = 1}^K {\left[ {H\left( {{q^{(k)}}} \right) - g} \right]}  + V\left( {1 - {q^{(K)}}} \right)} \right\}.
	\end{align}
	Applying a similar method as in the previous case, we can obtain
	\begin{align*}
	\frac{{\partial r({\omega ^l},\lambda )}}{{\partial \lambda }} =  - \frac{1}{{{P^{(L)}}}} < \frac{{\partial a({\omega ^l},\lambda )}}{{\partial \lambda }} =  - \frac{{{\omega ^l}}}{{{P^{(L)}}}} + 1.
	\end{align*}
	For $\omega^u$, since $K=\infty$, we can prove that $T({\omega ^u},{\omega ^l},{\omega ^u}) = \infty $ (See Lemma C4 in Appendix C). Hence $r(\omega^u,\lambda)$ is with the same form as $r(\omega^l,\lambda)$ given by \eqref{eq:LD3-2}. We thus obtain
	\begin{align*}
	\frac{{\partial r({\omega ^u},\lambda )}}{{\partial \lambda }} =  - \frac{1}{{{p^{(L)}}}} < \frac{{\partial a({\omega ^u},\lambda )}}{{\partial \lambda }} =  - \frac{{{\omega ^u}}}{{{p^{(L)}}}} + 1.
	\end{align*}
	
	Finally, if $L=\infty$ and $K=\infty$. Then $V(p)$ is given by \eqref{eq:LD3-1} and $V(1-q)$ is given by \eqref{eq:LD3-3}. For $\omega=\omega^l$ or $\omega^u$, we can verify that
	\begin{align*}
	\frac{{\partial r(\omega ,\lambda )}}{{\partial \lambda }} = 0 < \frac{{\partial a(\omega ,\lambda )}}{{\partial \lambda }} = \omega \frac{{\partial V(1 - q)}}{{\partial \lambda }} + 1 = 1.
	\end{align*}
	
	Putting the above results together, we can obtain the property stated in Lemma C3.
\end{IEEEproof}

\textit{\textbf{Lemma} C4:} Let $(\omega^l,\omega^u)$ denote the sampling region of the optimal policy of an oscillating bandit ($p+q>1$) with a service charge $\lambda$. If ${\omega ^u} > {\omega ^l} \ge {\omega ^ * }$ and $T(1 - q,{\omega ^l},{\omega ^u}) = \infty $, then $T(\omega^u,{\omega ^l},{\omega ^u}) = \infty $.
\begin{IEEEproof}
	If $p+q>1$ and ${\omega ^u} > {\omega ^l} \ge {\omega ^ * }$, according to Lemma 8, $T({\omega ^u},{\omega ^l},{\omega ^u}) = 2$ if ${\tau ^2}({\omega ^u}) > {\omega ^l}$; otherwise, $T({\omega ^u},{\omega ^l},{\omega ^u}) = \infty $. To prove this lemma, it suffices to show that $T({\omega ^u},{\omega ^l},{\omega ^u}) \neq 2$. Note that $1-q<\omega^*$.  As discussed in the proof of Lemma 8, if ${\omega ^u} > {\omega ^l} \ge {\omega ^ * }$, then  $T(1 - q,{\omega ^l},{\omega ^u}) = \infty $ holds in one of the following cases: (i) $\tau (1 - q) < {\omega ^l}$; (ii) there exists an integer $n$ such that ${\tau ^n}(1 - q) \ge {\omega ^u}$ and ${\tau ^{n + 2}}(1 - q) \le {\omega ^l}$. We show that $T({\omega ^u},{\omega ^l},{\omega ^u}) = 2$ contradicts these two conditions.
	
	First, if  $T({\omega ^u},{\omega ^l},{\omega ^u}) = 2$, then
	\begin{align*}
	{\omega ^l} &< {\tau ^2}({\omega ^u}) = p + (1 - p - q)\left[ {p + {\omega ^u}(1 - p - q)} \right]\\
	&< p + (1 - p - q)\left[ {p + (1 - p - q)} \right] = \tau (1 - q).
	\end{align*} 
	We thus get a contradiction with $\tau(1-q)<\omega^l$.
	
	Furthermore, if ${\tau ^2}({\omega ^u}) > {\omega ^l}$, then for any $n$ satisfying ${\tau ^n}(1 - q) \ge {\omega ^u}$, we have
	\begin{align*}
	{\tau ^{n + 2}}(1 - q) &= p + p(1 - p - q) + {\tau ^n}(1 - q){(1 - p - q)^2}\\
	&\ge p + p(1 - p - q) + {\omega ^u}{(1 - p - q)^2} \\
	& = {\tau ^2}({\omega ^u}) > {\omega ^l}.
	\end{align*}
	which contradicts the condition that ${\tau ^n}(1 - q) \ge {\omega ^u}$ and ${\tau ^{n + 2}}(1 - q) \le {\omega ^l}$. We thus conclude that $T({\omega ^u},{\omega ^l},{\omega ^u}) \ne 2$.
\end{IEEEproof}

\bibliographystyle{IEEEtran}
\bibliography{reference}

\begin{IEEEbiographynophoto}{Gongpu Chen}
	received his B.S. degree in automation engineering from University of Electronic Science and Technology of China (UESTC), Chengdu, China, in 2016, and the M.S. degree in Control Science and Engineering in Southeast University, Nanjing, China, in 2019. He is currently a PhD student in the Department of Information Engineering at The Chinese University of Hong Kong. His current research interests include cyber physical systems, wireless communications and networking, reinforcement learning, and multi-armed bandits. 
	
\end{IEEEbiographynophoto}

\begin{IEEEbiographynophoto}{Soung Chang Liew}
	received his  S.B., S.M., E.E., and Ph.D. degrees from the Massachusetts Institute of Technology. From 1984 to 1988, he was at the MIT Laboratory for Information and Decision Systems, where he investigated Fiber-Optic Communications Networks. From March 1988 to July 1993, he was at Bellcore (now Telcordia), where he engaged in Broadband Network Research. Since 1993. he has been a Professor at the Department of Information Engineering, the Chinese University of Hong Kong (CUHK).  Prof. Liew is currently a Co-Director of the Institute of Network Coding at CUHK. His research interests include wireless networks, Internet of Things, intelligent transport systems, Internet protocols, multimedia communications, and packet switch design. Prof. Liew is the recipient of the first Vice-Chancellor Exemplary Teaching Award in 2000 and the Research Excellence Award in 2013 at CUHK. Prof. Liew is a Fellow of IEEE, IET, HKIE, and Hong Kong Academy of Engineering Sciences.
	
\end{IEEEbiographynophoto}

\begin{IEEEbiographynophoto}{Yulin Shao}
	received his B.S. and M.S. degrees in Electronic and Communications Engineering from Xidian University (Hons.) in 2013 and 2016, and the Ph.D. degree in Information Engineering from The Chinese University of Hong Kong (CUHK) in 2020. He was a research assistant with the Institute of Network Coding (INC) from 2015 to 2016, a visiting scholar in Research Laboratory of Electronics at Massachusetts Institute of Technology (MIT) from 2018 to 2019, and a postdoctoral fellow in the Department of Information Engineering at CUHK from 2020 to 2021. He is currently a research associate in the Department of Electrical and Electronic Engineering at Imperial College London. His research interests include wireless communications and networking, signal
	processing, stochastic control, and machine learning.
	
\end{IEEEbiographynophoto}

\end{document}